\documentclass[modern]{aastex631}
\date{April 30, 2026}
\usepackage{graphicx} 
\usepackage{amssymb,amsmath}
\usepackage{booktabs}
\usepackage{ragged2e}
\newcommand{\customref}[2]{\hyperref[#1]{#2}}

\accepted{March 09, 2026}
\published{April 30, 2026}
\submitjournal{The Astrophysical Journal Supplement Series}

\shorttitle{NSDB: PCA/AI Analysis of Novae Spectra}
\shortauthors{Santos et al.}

\begin{document}

\title{The Nova Synthetic Data Base: a PCA/AI Analysis of Novae Synoptic Spectra}

\author[0009-0006-8325-4939]{Bruno C. Santos}
\author[0000-0002-6040-0458]{Marcos P. Diaz}
\author[0000-0002-2035-5277]{Larissa Takeda}
\affiliation{Instituto de Astronomia, Geofísica e Ciências Atmosféricas da Universidade de São Paulo (IAG-USP) \\ Rua do Matão, 1226, Cidade Universitária \\ 05508-090 São Paulo, SP Brazil}

\begin{abstract}

The Nova Synthetic Data Base (NSDB) is presented as the first publicly available database of synthetic spectra for classical nova shells, spanning an unprecedented range of physical parameters (e.g., ejecta mass, chemical composition, temperature, and luminosity of the white dwarf) at several post-eruption ages. Generated using detailed 3D photoionization models, this homogeneous database enables a systematic exploration of spectral features in novae. In this work, we introduce a PCA/AI-based framework to derive time-dependent proxies for retrieving the physical properties of novae from limited spectral data. By analyzing the correlations between the eigenspectra and the grid's variables, a reduced set of diagnostic spectral lines is derived, paving the way for robust multi-regressor machine-learning algorithms with a minimal effort observational set. The prediction capability of the method is high and robust to data noise. The results establish a proof of concept for the use of model grids combined with physically controlled AI as a tool to interpret novae observations in the context of the large number of events expected from future wide area surveys.

\end{abstract}


\section{Introduction} \label{sec:Introduction}

Cataclysmic variables are binary systems in which the secondary component transfers mass to a white dwarf. This interaction between stellar components leads to unique observational signatures, one of which being among the most energetic stellar phenomena in the Universe: classical novae. The accretion of material onto the degenerate primary builds up a hydrogen-rich layer, eventually triggering a thermonuclear runaway (TNR). This results in a recurrent eruption, ejecting shells that expand at velocities of $\sim10^3$ km\,s$^{-1}$ and a range of mass loss of $10^{-7}-10^{-4}$ M$_\odot$ per event. Chemical signatures from both the primary and secondary, along with the products from TNR nucleosynthesis, are returned to the ISM \citep{Warner1995}. Beyond being theorized as potential progenitors of Type Ia Supernovae, these systems play an important role in synthesizing $^7$Li and rare isotopes \citep{Starrfield1978}. Their study is important for understanding the details of galactic chemical evolution \citep{Borisov2024} and for constraining the Single Degenerate progenitor channel \citep{Maoz2014} of Type Ia Supernovae. Improving our knowledge of these progenitors is, in turn, a key step toward further refining the most widely used method for determining cosmological distances \citep{Noble1,Noble2}.

Observed nova spectra diverge significantly from those depicted by one-dimensional photoionization models with acceptable physical parameters. The simultaneous presence of spectral lines of neutral elements and highly ionized species at several post-eruption ages \citep{Williams1992}, along with the large optical depth of lines like [O \textsc{i}], can only be explained by high-density mass clumps embedded in the diffuse plasma during the nebular phase \citep{Williams1994,Diaz_2010}. Spatially resolved observations of nova shells \citep{Slavin1995,Moraes2009,Diaz2018} clearly show a complex structure and the presence of mass clumps in classical nova ejecta.

Various three-dimensional modeling efforts of ionized nebulae have been attempted at high computational cost \citep[e.g.][]{Gruenwald1997,Ercolano2003}. In this context, a three-dimensional approach for modeling the nebular anisotropic medium was proposed in the photoionization code RAINY3D \citep[][see also \customref{sec:Methodology-Nova Synthetic Data Base}{Section \ref{sec:Methodology-Nova Synthetic Data Base}}]{Moraes&Diaz2011}. Previous works have demonstrated the capability of this approach to replicate the physical environment of classical novae in the nebular phase. \citet{Takeda2018} presented a dive into the slow novae \textit{V723 Cas} photoionization structure, showing that more realistic anisotropic mass and radiation field distributions are critical for reproducing the system’s complex features, unlike simplistic 1D approaches.

The forthcoming increase in public spectroscopic novae data (e.g., VRO, ZTF-BTS) will soon overwhelm our efforts in individual modeling, while opening unprecedented nova population and evolution studies if basic parameters can be obtained directly from the observed spectrum. In this context, the qualitative classification scheme of \citet{Williams1991} - based on emission line evolution and observable properties - pioneered the systematic observational framework of novae categorization and correlation of spectral properties with global parameters, i.e. mean mass gas density and ionizing radiation field.  Machine learning methods trained on synthetic model grids offer a promising follow-up alternative, but only if spectral features can be unambiguously and quantitatively mapped to specific physical parameters.

In this paper, we present a \customref{sec:Methodology-PCA Decomposition}{Principal Component Analysis} (PCA) of the large NSDB 3D synthetic spectral grid for classical novae: the \customref{sec:Methodology-Nova Synthetic Data Base}{\textit{Nova Synthetic Data Base}}. This grid comprises 1439 models (875 for the CO nova grid, 564 for the Ne enhanced nova grid) of anisotropic shells, with condensed mass clumps added to a symmetric power-law background, spanning a wide range of physical parameters. By identifying correlations between the eigenspectra and the physical parameters, we were able to identify sets of diagnostic spectral lines that act as time-dependent proxies for the shell and central source properties. This framework not only enables rapid parameter ranging from incomplete observational data but also sets the stage for AI-driven analyzes. PCA preliminary analysis favors a closer physical insight into the AI methods and their application to next-generation nova synoptic data, without the need for single-object photoionization modeling.

\section{Methodology} \label{sec:Methodology}

\subsection{The Nova Synthetic Data Base} \label{sec:Methodology-Nova Synthetic Data Base}

The \textit{Nova Synthetic Data Base}\label{def:Nova Synthetic Data Base} is a synthetic spectral database of nova shells at several evolutionary stages.  The grid is defined by $6$ basic free parameters, which we believe represent the most relevant physical properties of the classical novae photoionization scenario, namely the post-Nova age, ejected mass, condensed mass-fraction, C,N,O abundance, luminosity, and temperature of the central source. Two separate similar grids are available, one of them is calculated using typical neon nova enhanced Ne abundance.

The C, N, and O content is varied with fixed fractions by $\pm1$ dex around averaged nova values, mostly taken from \citet{Gehrz1998} compilation ($\log{N_{C,N,O}/N_H} = -2.06,-1.847,-1.77$). Solar abundances \citep{Asplund2009} were assumed for all elements other than He, C, N, O, and Ne for the neon nova grid. Nova ejected masses ranging from $8 \times 10^{-6}$ to $10^{-4}$ M$_\odot$ are intended to cover the most commonly observed values in classical novae regime. The post-eruption ages between $80 - 2560$ days focus on the nebular phase of fast ejected shells. This initial version of the model grid assumes a constant expansion velocity profile, with effective velocities of $1500$, $950$, and $400$ $km \ s^{-1}$ as maximum, average, and minimum velocities, completely coupling the inner and outer radii as well as the mean density of the simulated shell with the post-eruption age. Consequently, slower or even faster novae may find their place in the grid by scaling their post-eruption age. The shell is assumed to have a constant velocity expansion for the radiative transfer. The central source properties also aim to cover values often reported in the literature for classical novae. Its SED is simulated from high-gravity hot NLTE spectra by \citep{Rauch2003}.  More details are given in \customref{tab:input_parameters}{Table 1}.

\begin{deluxetable*}{lccc}
\tabletypesize{\normalsize}
\tablewidth{1.0\textwidth} 
\tablecaption{NSDB Grid Input Parameters \label{tab:input_parameters}}
\tablehead{
\colhead{Parameters} & 
\colhead{Minimum Value} & 
\colhead{Maximum Value} & 
\colhead{Number of Values}
} 
\startdata
Age & $80$ \textit{days} & $2560$ \textit{days} & 6 \\
Ejected Mass (M) & \(8 \times 10^{-6} \, M_\odot\) & \(1 \times 10^{-4} M_\odot\) & 3 \\
Condensed fraction (fc) & 0.2 & 0.8 & 4 \\
Luminosity (log(L)) & \(36.5\) & \(38.5\) & 3 \\
Temperature (T) & \(1 \times 10^5 \, \mathrm{K}\) & \(3.5 \times 10^5 \, \mathrm{K}\) & 3 \\
C,N,O abundance ($\log{N_x/N_H}$) & -3.06,-2.847,-2.77 & -1.06,-0.847,-0.77 & 2 \\
\enddata
\tablecomments{This table summarizes the parameter ranges and the number of values sampled for each parameter, with luminosity in CGS units. Models of age 2560 days do not include the highest central source luminosity value. The CO Grid adopts solar Ne abundance, whereas the Neon Grid adopts $\log{N_{Ne}/N_H} = -2.07$.}
\end{deluxetable*}

Paving the way for a systematic exploration of the spectral signatures of condensations, a range of $20\%$ to $80\%$ condensed mass-fraction relative to total mass (fc) is assumed. Less changes in the resulting spectrum are seen beyond these limits because higher values lead to shells where the gaussian condensation tails blend with the diffuse power-law component and with each other, whereas lower values result in spectra dominated by emission from the symmetric component. There are very few constraints on the mass fraction in the literature. The work by \citet{Abraham2024} indicates fc $\sim0.4$ for the shell around nova V5668 Sgr. The condensation radius (FWHM) is a fixed fraction (1/8) of the shell outer radius, while their peak density contrast relative to the diffuse background is 20; values inspired by the ALMA imaging of V5668 Sgr \citep{Diaz2018}. The symmetric radial power-law background has an index of -1.4 for CO novae and -1.0 for neon novae. These values were roughly estimated from the H$\alpha$ emissivity fitting and deprojection of a few resolved shells. The integrated spectrum depends weakly on the power-law index, given that the inner radius of the shell is a significant fraction of the shell thickness.

The models were constructed using the anisotropic photoionization code RAINY3D\label{def:RAINY3D} \citep{Moraes&Diaz2011}. The code builds an expanding shell with two mass distribution components: a radially symmetric power-law and gas clumps with Gaussian contrast density profiles and sizes, both of which may be governed by power-law probability functions. However, they were predefined in the present grids instead. Random clump positions, embedded in the spherically symmetric background, are assumed. The condensed mass-fraction parameter (fc) is more appropriate than filling or covering factors when comparing clumpy models with different density contrasts. Once the anisotropic shell is defined, the photoionization code CLOUDY \citep{Ferland2013} is called as a subroutine to calculate the 1D photoionization and thermal equilibrium, as well as the radial radiative transfer of the white dwarf's and diffuse radiation fields. RAINY3D has a feature to include an accretion disk contribution to the radiation field, parameterized by the accretion rate and the white dwarf mass; a feature that was not used in the presented results.

As an output of every model, RAINY3D provides local and integrated line fluxes, accounting for the entire shell, of $241$ selected spectral lines of different elements at several ionization stages, which compose possible nova remnant spectra. Calculated line fluxes correspond to observed values at a distance of 1.0 kpc in the absence of extinction. The line list was selected on the basis of Cloudy's atomic lines database and on the lines that are commonly found in nova visible and NIR spectra available in the literature. A rest wavelength range from $\lambda3200$ \AA\ to $\lambda24,800$ \AA\ (from near-UV to NIR) was chosen; the full spectral line selection can be found at \customref{Appendix:RAINY3D Spectral Lines Output}{Appendix \ref{Appendix:RAINY3D Spectral Lines Output}}.

The grid was entirely computed using the Santos Dumont (SDumont\label{def:SDumont}) supercomputer facility at the Laboratório Nacional de Computação Científica (LNCC), with a hybrid CPU architecture and a peak performance of $5.1$ PetaFLOPs.

Ultimately, $875$ novae remnants models compose the final CO novae grid, described by $241$ total line fluxes per model, covering the optical to near-infrared wavelengths. This grid is a homogeneous dataset where individual model structures, including local emissivities and physical conditions, can be retrieved. An enhanced neon grid was also computed, aiming to model the neon nova spectral evolution. The NSDB data are publicly available at the IAG-USP model repository: \url{http://specmodels.iag.usp.br}.

\subsection{Line Fluxes and Variable Spaces} \label{sec:Methodology-Lines Fluxes and Variables Spaces}

Before detailing the method of analysis of this extensive data set, we first define the variable spaces that will be treated: The most direct variables that map the world of models are the input variables of RAINY3D (listed in \customref{tab:input_parameters}{Table 1}). This mapping is called \textit{Parameter-Space}\label{def:Parameter-Space}, which describes the $875$ CO models through the $6$ input parameters of the code used to create the grid.

Another way to represent the computed shells is through the resulting integrated line flux sets. Each spectrum was normalized by its H$\beta$ ($\lambda$$\lambda4861.33$ \AA) value, resulting in the \textit{Spectral-Space}\label{def:Spectral-Space}. This space maps the world of models through $240$ selected line fluxes relative to H$\beta$. These lines are or may be observed in nova ejecta at all ages. In this space, the eigenvalues and eigenspectra decomposition will be performed in order to find a representation base, eigenspectra, which better describes the variance of line fluxes within each age subset.

Aiming to construct the representation base, the $100$ brightest spectral lines were extracted from our previously selected list in order to minimize noise from stochastic variable realization and statistical fluctuations introduced by RAINY3D's pseudo-random processes while avoiding, as much as possible, the use of lines that cannot be observed at moderate signal-to-noise ratios (despite their potential diagnostic power). Therefore, this selection criterion prioritizes lines with the highest flux ratios relative to H$\beta$.

The dataset subjected to this selection is the epoch-wise spectral space, or the $100$ main spectral lines that better describe each post-eruption age. This will be the \textit{Epoch-Selection}\label{def:Epoch-Selection}, built for each of the $6$ post-eruption ages modeled. All $6$ spectral line selections are presented in \customref{Appendix:Epoch-Selections}{Appendix \ref{Appendix:Epoch-Selections}}.

\subsection{PCA Decomposition} \label{sec:Methodology-PCA Decomposition}

Dealing with computed models (i.e. without observational biases, incomplete sampling, stochastic noise, and extinction correction uncertainties), it is recommended to perform the eigenvalues-eigenvectors decomposition directly on a homogeneous \textit{data correlation matrix}\label{def:data correlation matrix} \citep{Murtagh2012}: the line flux sample is centralized and scaled to unit variance (standardization\label{def:standardization}/Z-score\label{def:Z-score}) for each Spectral-Space's variable (selected spectral lines). In addition, the sample is normalized by the number of models considered. This creates a space where the distance between variables is proportional to their correlation, preserving the weight of weak line relative variance. Once the decomposition of the data correlation matrix is performed, the eigenspectra are ranked by their cumulative explained variance. A $90\%$ threshold is adopted to ensure the capture of the primary physical-related signal while effectively filtering out noise and most uncertainties intrinsic to integrated line luminosity from observations.

It is worth mentioning that we expect the decomposition to retrieve a number of eigenvectors similar to or greater than the hidden variables of the Parameter-Space ($5$ per age). A smaller number of axes would mean that some physical variable used to construct the model's grid is degenerate or fully correlated with another variable.

With the decomposition method at hand, it may be applied to various synthetic spectra datasets along with their respective line selections. As a result, $6$ eigenspectra were obtained, with the number of eigenvectors per base varying from $5$ to $9$ for the Epoch-Selections. Each eigenvector in these eigenspectra \label{def:eigenspectra} is a vector with $100$ components, which represents, in its absolute value, the relative contribution of the corresponding spectral line to that eigenvector.

Mathematically, these eigenspectra define a new orthogonal basis that maximizes the variance of the model ensemble, providing vectors (or axes) that better map and describe the dataset.

\subsection{Variables Correlation \& Diagnostic Lines} \label{sec:Methodology-Variables Correlation & Diagnostic Spectral Lines}

To explore possible correlations between the PCA-derived basis and the fundamental novae physical parameters, the standardized data are projected onto the new basis, and the \textit{variables correlation matrix} is computed\label{def:variables correlation matrix}: This matrix contains Pearson correlation coefficients between the \customref{def:eigenspectra}{eigenspectra} obtained and the \customref{def:Parameter-Space}{Parameter-Space} input variables, calculated using the \textit{corrwith} \label{def:corrwith()} method from the Pandas library \citep{McKinney2010, pandas_2_1_4}.

Searching for significant correlations, we were able to predict, among the $100$ brightest expected spectral lines in the optical and near-infrared ranges, the most useful diagnostic lines for each age. Diagnostic lines show a greater relative contribution to eigenspectra and correlation with physical parameters. Defining a minimum absolute correlation of $0.25$ with any physical variable, the diagnostic significance of a spectral line for the diagnosis of that variable is calculated using the following formula:

\begin{equation}
    I_{i,k} = \sum_j ES_i^j \times |corr(ES^j,V_k)| \times P_j \ \text{;}
    \label{eq:I_{i,k}}
\end{equation}

\noindent where the importance of the spectral line $i$ for the diagnosis of the variable $k$, $I_{i,k}$, is given by the sum of its contribution to the $j$-th eigenspectra, $ES_i^j$, weighted by the absolute correlation of that eigenspectra with the variable under analysis and by the percentage of total variance explained by the $j$-th eigenspectra $P_j$. The last weighting term is justified by both preserving the quality ranking in the description of generic novae spectra and reducing the influence of spurious correlations and noise in the analysis, which is critical when an application to observed spectra is intended.

\subsection{Machine Learning Application} \label{sec:Methodology-Machine Learning Application}

Aiming to establish robust diagnostics of nova spectra, an AI-based algorithm should be able to inversely map spectral features to physical parameters using only the most relevant diagnostic lines derived from PCA analysis. For this reverse prediction task, we implemented a \textit{Random Forest} \citep{Breiman2001} regressor \label{def:Random_Forest} based on three critical properties:

\begin{itemize}
    \item \textit{Bootstrap aggregation:} each tree trains on randomized data subsets (with replacement), maximizing information extraction from a limited model grid.
    \item \textit{Double randomness} mechanism - random data subsets per tree and random feature subsets at node splits - minimizes tree correlations and, therefore, mitigates \textit{overfitting}.
    \item This architecture accepts \textit{transfer learning} techniques for future applications to observed data. 
\end{itemize}

Considering that tree-based regressors perform poorly in extrapolating beyond the training range, our model's grid was designed to cover the most common physical parameter space of classical novae.

To ensure reliable diagnostics, we require a homogeneous set in the Parameter-Space. For this purpose, and because of the lack of convergence of some models in the two earliest epochs, the subsets with age in the range of $320$ - $2560$ days were selected.

For the hyperparameter optimization, different regressors were evaluated using the \textit{mean absolute error} (MAE) metric, exploring eight optimization parameters of the Random Forest Regressor ensemble, as available in the Scikit-learn library \citep{pedregosa2011scikit}: the number of trees (with a minimum number of $100$ to ensure at least one usage of each spectral line selected in a random-walk-like selection process), maximum features per tree, maximum tree depth, minimum samples per leaf node, minimum samples for node splitting, and three other secondary parameters accounting for the split loss function, minimum impurity decrease per division, and pruning complexity parameter. The search employed 5-fold cross-validation with randomized partitioning, following established best practices for astrophysical machine learning applications \citep{ivezic2019}.

After establishing the best model, we evaluate its performance using two complementary approaches:

\begin{enumerate}
    \item[(1)] The \textit{out-of-bag} (OOB) score - an intrinsic Random Forest metric based on bootstrap aggregation, where each tree is evaluated on unseen data from its training set. This provides an estimate of the model's generalization without requiring a separate validation data set.
    \item[(2)] Parameter-Space specific metrics, including MAE and $R^2$ \textit{coefficient of (multiple) determination} \label{def:R2_coefficient} - interpreted as the fraction of a variable variance explained by the model and are widely used as a first adequacy indicator for any multiple regression model \citep[Cap. 13]{Devore2011}.
\end{enumerate}

\noindent These metrics were computed epoch-wise to account for evolutionary dependencies in our database.

The optimization and evaluation are repeated for a decreasing number of diagnostic lines, building regressor models for the top $n$ most relevant diagnostic lines according to the PCA analysis described above. The smaller $n$, while achieving satisfactory retrieval of physical parameters, represents the minimum observational effort and greater dimensionality reduction. \label{def:Top_n_Regressor}

\subsection{Simulating Observations} \label{sec:Methodology-Simulating Observations}

As a final test of the method, Gaussian noise is applied to the Spectral-Space, simulating observational data of Novae in preparation for future applications of the framework constructed here. One-sigma uncertainties are applied to the simulated flux,  from $0\%$ up to $20\%$ of the flux, and a hundred realizations are made to achieve statistical relevance at each uncertainty level. The final evaluation metrics are then computed for each bootstrap.

The performance evaluation against noisy data follows specific metrics\label{def:MAPE_and_MALE} to account for different ranges, sampling, and performance differences in Parameter-Space. The absolute relative error is derived for the linear grid variables, namely the ionizing source temperature and the condensed mass fraction of the envelope, while the absolute logarithmic error is calculated for logarithmic grid variables such as the ejected mass, C,N,O abundance, and central source luminosity. The median central value is computed for each physical parameter. The \textit{MAPE} and \textit{MALE} define an uncertainty-like metric, in percentage and in dex, for our regressor model. They grade the retrieval of Nova physical parameters from limited spectroscopic data.

\section{Results \& Discussion} \label{sec:Results}

Firstly, we present the $6$ eigenspectra decomposition using the \customref{def:Epoch-Selection}{Epoch-Selection} in \customref{tab:eigenspectra_Epoch-Selection}{Table \ref{tab:eigenspectra_Epoch-Selection}}. Notably, the model spaces for the two earliest post-eruption epochs (80 and 160 days) and the latest (2560 days) are well described by only five and six eigenspectra, respectively – 3 or 4 fewer than required for intermediate epochs. This is directly linked to the reduced physical parameter space of these subsets. At the early stages, unstable models were disregarded from the model grid. For the latest epoch, the highest white dwarf luminosity is physically excluded, as the central source has possibly cooled by that time. These limitations create reduced ranges at these epochs. Consequently, the number of models drops from $217$ to $37$ at 80 days, $46$ at 160 days, and $144$ at 2560 days. This reduction in input parameters within the \customref{def:Parameter-Space}{Parameter-Space} enhances the correlations between the eigenspectra and these critical variables, as detailed in \customref{tab:Maximum_Correlation_Coefficients}{Table \ref{tab:Maximum_Correlation_Coefficients}}.

\begin{deluxetable}{cccc}
\tabletypesize{\normalsize}
\tablewidth{1.0\textwidth} 
\tablecaption{Eigenspectra Decomposition \label{tab:eigenspectra_Epoch-Selection}}
\tablehead{
\colhead{\shortstack{Age (\textit{days})}} & 
\colhead{\shortstack{no. of \\ Eigenspectra}} & 
\colhead{\shortstack{Maximum \\ Variance Explained}} & 
\colhead{\shortstack{Minimum \\ Variance Explained}} 
} 
\startdata
$80$ & $5$ & $50.0\%$ & $3.9\%$  \\
$160$ & $5$ & $43.1\%$ & $4.9\%$  \\
$320$ & $9$ & $25.9\%$ & $2.5\%$  \\
$640$ & $9$ & $28.8\%$ & $2.4\%$  \\
$1280$ & $8$ & $30.6\%$ & $2.5\%$  \\
$2560$ & $6$ & $36.8\%$ & $4.9\%$  \\
\enddata
\tablecomments{Number of eigenvectors necessary to achieve $90\%$ of Spectral-Space' variance and their maximum and minimum eigenvectors' explained variance for each age using the Epoch-Selection lines.}
\end{deluxetable}

A global line selection (accounting for all post-Nova ages) was attempted, requiring more eigenvectors to reach the $90$ variance threshold without bringing any gain in the correlations between the input parameters (including post-eruption age) and these eigenvectors. This reflects the complexity of spectral evolution in novae and suggests that post-eruption age, which defines size scales and densities, is an essential observable that defines the diagnostic framework. We therefore focus on the Epoch-Selection, which achieves significant dimensionality reduction – compressing the \customref{def:Spectral-Space}{Spectral-Space} from $100$ spectral lines to $5$–$9$ eigenspectra while preserving diagnostic power.

\subsection{Eigenspectra \& Correlations} \label{sec:Results-Eigenspectra & Correlations}

As outlined in \customref{sec:Methodology-PCA Decomposition}{Section \ref{sec:Methodology-PCA Decomposition}}, each eigenspectra composing the basis is a vector whose components represent the absolute contribution of the corresponding spectral line to that eigenvector, given the post-eruption age and the emission line sample. These contributions arise from the variance (in standard deviation units) of each spectral line ratio across age-specific model sets and their diagnostic power in discriminating models with varying physical parameters.

\begin{figure}
    \centering
    \includegraphics[width=1.0\textwidth]{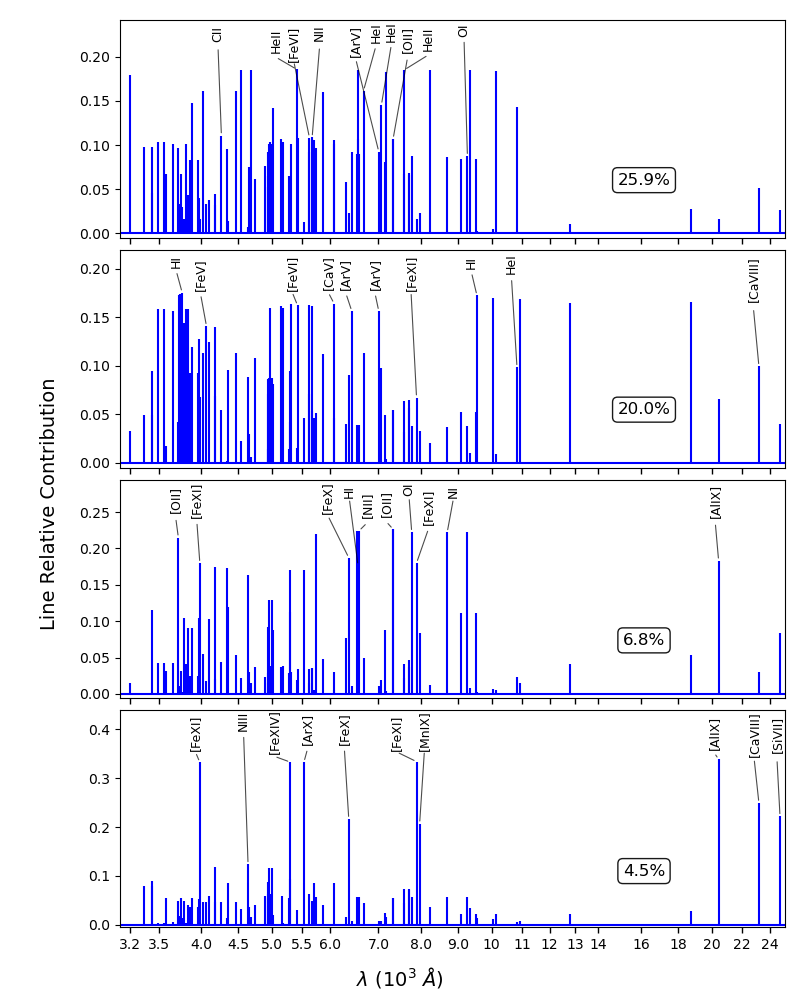}
    \vspace{-10pt}
    \caption{Line relative contribution to variance of the, from top to bottom, higher representative eigenspectra $1$ and $2$, and lower eigenspectra $5$ and $7$, at $320$ days post-eruption using the Epoch-Selection, with their respective variance explained as percentages. The black tags label the top line ratio of the 5 most important ions in the visible ($<7000$ \AA) and the 5 most important ions in the NIR ($>7000$ \AA), ranked by their contribution in each case. Wavelength axis in log scale.}
    \label{fig:Eigenspectra_320d}
\end{figure}

\customref{fig:Eigenspectra_320d}{Figure \ref{fig:Eigenspectra_320d}} illustrates representative eigenspectra, plotted as delta functions at their respective wavelengths. Both figures show general features present in all post-eruption ages:

\begin{enumerate}
    \item Dominant eigenspectra (those explaining the most variance - represented by \customref{fig:Eigenspectra_320d}{Figure \ref{fig:Eigenspectra_320d}} top panels) are overwhelmingly dominated by He lines (neutral and ionized), with consistent but secondary contributions from H lines. Highly ionized metals (e.g., Ca $\textsc{v}$, Ar $\textsc{v}$, Fe $\textsc{vii}$) become increasingly prominent in these components beyond 640 days. The distribution between optical and NIR lines varies with age, showing no trend or clustering.
    \item Minor eigenspectra primarily feature optical-region lines from highly ionized metals (e.g., Fe $\textsc{vi}$, Fe $\textsc{vii}$, N $\textsc{iv}$, Ne $\textsc{v}$, O $\textsc{iv}$, Ar $\textsc{v}$), though NIR features (e.g., Al $\textsc{ix}$) appear in specific eigenvectors. Their significance seems age-dependent, peaking at intermediate evolutionary stages (160–1280 days).
\end{enumerate}

This scenario is aligned with the expected nebular line formation \citep{Osterbrock2006} and PCA’s discrimination.  The overall ionization structure of the envelope can be found through the analyzes of H, He $\textsc{i}$ and He $\textsc{ii}$ lines, giving rise to their large contribution in the main eigenspectra - The migration of ionized metals (e.g. Ar $\textsc{v}$/Fe $\textsc{vii}$) into these dominant components at late ages reflects a shift in the ionization regime of the remnants. The contribution of ionized metal lines' in minor eigenspectra is a consequence of their diagnostic power between similar models, enabling fine mapping within Spectral-Space.

With the eigenspectra at hand, the correlation between these newly derived axes and the input variables of the \customref{def:Parameter-Space}{Parameter-Space} - the critical physical variables for the construction of the \customref{def:Nova Synthetic Data Base}{Nova Synthetic Data Base} - can be pursued.

\customref{tab:Maximum_Correlation_Coefficients}{Table \ref{tab:Maximum_Correlation_Coefficients}} compiles peak correlations between parameters and age-specific eigenspectra using the Epoch-Selection. All parameters, except fc (mass fraction in condensations), show significant correlations ($\gtrsim 0.3$) across epochs. The low correlation values for fc across all epochs are noteworthy. While a purely statistical approach might suggest a negligible role for this parameter, the physical significance of condensations has been extensively demonstrated \citep[e.g.,][]{Williams1994, Diaz_2010}. Moreover, the application of the RAINY3D approach in previous spectral syntheses (e.g. \citet{Takeda2018}) supports the view that fc is a critical, albeit complex, driver of the spectral evolution in nova remnants. The low correlation indicates that the spectral response to high density clumps cannot be simplified into few linear components, unlike the more straightforward responses to parameters like effective temperature, luminosity, or ejected mass.

\begin{deluxetable*}{ccccccc}
\tabletypesize{\normalsize}
\tablewidth{1.0\textwidth} 
\tablecaption{Maximum Correlation Coefficients for Epoch-Selection \label{tab:Maximum_Correlation_Coefficients}}
\tablehead{
\colhead{Parameter} & 
\colhead{$80$ \textit{days}} & 
\colhead{$160$ \textit{days}} & 
\colhead{$320$ \textit{days}} &
\colhead{$640$ \textit{days}} &
\colhead{$1280$ \textit{days}} &
\colhead{$2569$ \textit{days}} 
} 
\startdata
M & $0.47$ & $0.40$ & $0.50$ & $0.36$ & $0.27$ & $0.43$ \\
fc & $0.26$ & $0.10$ & $0.04$ & $0.05$ & $0.04$ & $0.06$ \\
L & $0.60$ & $0.68$ & $0.46$ & $0.38$ & $0.38$ & $0.44$ \\
T & $0.70$ & $0.65$ & $0.78$ & $0.75$ & $0.70$ & $0.89$ \\
C,N,O abundance & -/- & -/- & $0.64$ & $0.65$ & $0.62$ & $0.67$ \\
\enddata
\tablecomments{Maximum (absolute) correlation coefficients between the variables in the Parameter-Space and the eigenspectra of each age. Models of age $80$ and $160$ days does not include C,N,O abundance as a variable.}
\end{deluxetable*}

\subsection{Novae Diagnostic Spectral Lines} \label{sec:Results-Novae Diagnostic Spectral Lines}

The significant correlations between Parameter-Space variables and eigenspectra for each age (see \customref{tab:Maximum_Correlation_Coefficients}{Table \ref{tab:Maximum_Correlation_Coefficients}}) enable the identification of the main diagnostic spectral lines for each parameter. Using \customref{eq:I_{i,k}}{Equation (\ref{eq:I_{i,k}})}, we calculate the percentage relevance of spectral lines for each variable; below we have compiled the \textit{top $10$ diagnostic spectral lines}\label{def:Top 10 Diagnostic Lines} within the Spectral-Space epoch-wise:

\begin{deluxetable}{lccc}
\tablecaption{Top 10 Diagnostic Lines: Early and Late Post-Eruption Ages\label{tab:top_lines_combined}}
\tablehead{
\colhead{Rank} & 
\colhead{80 days} & 
\colhead{160 days} & 
\colhead{320 days}
}
\startdata
1 & H $\textsc{i}$ ($\lambda$$\lambda$9545.93) & He $\textsc{i}$ ($\lambda$$\lambda$6678.15) & He $\textsc{ii}$ ($\lambda$$\lambda$5411.37) \\
2 & He $\textsc{i}$ ($\lambda$$\lambda$7065.22) & He $\textsc{i}$ ($\lambda$$\lambda$5015.68) & He $\textsc{ii}$ ($\lambda$$\lambda$10123.3) \\
3 & H $\textsc{i}$ ($\lambda$$\lambda$19445.4) & He $\textsc{ii}$ ($\lambda$$\lambda$3923.37) & H $\textsc{i}$ ($\lambda$$\lambda$3770.63) \\
4 & H $\textsc{i}$ ($\lambda$$\lambda$9014.87) & He $\textsc{i}$ ($\lambda$$\lambda$5875.64) & [Fe $\textsc{vii}$] ($\lambda$3586.32) \\
5 & He $\textsc{i}$ ($\lambda$$\lambda$4713.03) & [Ne $\textsc{iii}$] ($\lambda$3868.76) & He $\textsc{ii}$ ($\lambda$$\lambda$6559.91) \\
6 & He $\textsc{i}$ ($\lambda$$\lambda$5875.64) & He $\textsc{ii}$ ($\lambda$$\lambda$3857.96) & H $\textsc{i}$ ($\lambda$$\lambda$3750.15) \\
7 & [Ne $\textsc{iii}$] ($\lambda$3868.76) & He $\textsc{i}$ ($\lambda$$\lambda$4921.93) & [Fe $\textsc{vii}$] ($\lambda$3758.92) \\
8 & N $\textsc{v}$ ($\lambda$$\lambda$4945.0) & [Ne $\textsc{iii}$] ($\lambda$3967.47) & He $\textsc{ii}$ ($\lambda$$\lambda$9344.62) \\
9 & He $\textsc{i}$ ($\lambda$$\lambda$6678.15) & He $\textsc{ii}$ ($\lambda$$\lambda$6682.98) & H $\textsc{i}$ ($\lambda$$\lambda$3734.36) \\
10 & [Ne $\textsc{iii}$] ($\lambda$3967.47) & He $\textsc{i}$ ($\lambda$$\lambda$4471.49) & [Fe $\textsc{vii}$] ($\lambda$5276.38) \\
\midrule
\multicolumn{1}{l}{Rank} & \multicolumn{1}{c}{640 days} & \multicolumn{1}{c}{1280 days} & \multicolumn{1}{c}{2560 days} \\
\midrule
1 & He $\textsc{ii}$ ($\lambda$$\lambda$3923.37) & He $\textsc{ii}$ ($\lambda$$\lambda$10123.3) & [Fe $\textsc{vii}$] ($\lambda$4942.48) \\
2 & He $\textsc{ii}$ ($\lambda$$\lambda$4541.46) & H $\textsc{i}$ ($\lambda$$\lambda$4340.46) & He $\textsc{ii}$ ($\lambda$$\lambda$3923.37) \\
3 & He $\textsc{i}$ ($\lambda$$\lambda$6678.15) & H $\textsc{i}$ ($\lambda$$\lambda$10938.0) & [Ar $\textsc{v}$] ($\lambda$6435.12) \\
4 & C $\textsc{iv}$ ($\lambda$$\lambda$4659.0) & C $\textsc{iv}$ ($\lambda$$\lambda$4659.0) & [Fe $\textsc{vii}$] ($\lambda$5276.38) \\
5 & He $\textsc{ii}$ ($\lambda$$\lambda$3857.96) & He $\textsc{ii}$ ($\lambda$$\lambda$6559.91) & He $\textsc{ii}$ ($\lambda$$\lambda$6682.98) \\
6 & H $\textsc{i}$ ($\lambda$$\lambda$19445.4) & H $\textsc{i}$ ($\lambda$$\lambda$6562.81) & [Ar $\textsc{v}$] ($\lambda$7005.83) \\
7 & He $\textsc{i}$ ($\lambda$$\lambda$4471.49) & He $\textsc{ii}$ ($\lambda$$\lambda$5411.37) & [Fe $\textsc{vii}$] ($\lambda$4988.55) \\
8 & [Ar $\textsc{v}$] ($\lambda$6435.12) & [Ar $\textsc{v}$] ($\lambda$7005.83) & He $\textsc{ii}$ ($\lambda$$\lambda$10123.3) \\
9 & He $\textsc{ii}$ ($\lambda$$\lambda$6682.98) & He $\textsc{ii}$ ($\lambda$$\lambda$4685.64) & [Ca $\textsc{v}$] ($\lambda$5309.11) \\
10 & H $\textsc{i}$ ($\lambda$$\lambda$21655.1) & N $\textsc{iii}$ ($\lambda$$\lambda$4641.0) & [Fe $\textsc{vii}$] ($\lambda$3586.32) \\
\enddata
\tablecomments{Top diagnostic lines ranked by importance for post-eruption ages within the NSDB. Upper panel shows early ages with transition from hydrogen/helium dominated to higher ionization species. Lower panel shows late ages with increased dominance of high-ionization species and iron. An extended version of this table is available at the database website.}
\end{deluxetable}

For better visualization, we group the $241$ spectral lines of \customref{def:RAINY3D}{RAINY3D} into $8$ categories and compute their percentage of importance: H $\textsc{i}$, He $\textsc{i}$ and He $\textsc{ii}$; Neutral C,N,O lines; Coronal lines including [Fe $\textsc{xiv}$] ($\lambda5303.01$), [Ar $\textsc{x}$] ($\lambda5534.02$), [Ca $\textsc{viii}$] ($\lambda23211.7$), among others; Classic Diagnostic [O $\textsc{iii}$], [N $\textsc{ii}$], and [S $\textsc{ii}$] as electron temperature and density diagnostic lines; and Other Forbidden, accounting for forbidden lines not already accounted for.

\begin{deluxetable}{lcccccc}
\tablecaption{Percentage Contribution of Main Diagnostic Lines \label{tab:Diagnostic_Lines}}
{\footnotesize
\tablehead{
  \colhead{Line Type} & \colhead{80d (\%)} & \colhead{160d (\%)} & \colhead{320d (\%)} & \colhead{640d (\%)} & \colhead{1280d (\%)} & \colhead{2560d (\%)}
}
\startdata
\multicolumn{7}{c}{\textit{Mass}} \\
H $\textsc{i}$             & 25.7 & 19.2 &  8.2 & 21.7 & 26.2 & 16.5 \\
He $\textsc{i}$            & 15.4 & 20.1 & 11.9 &  6.6 &  5.3 &  8.2 \\
He $\textsc{ii}$           & 15.2 & 21.9 & 18.8 & 22.7 & 27.7 &  7.8 \\
Neutral C,N,O              &  0.0 &  0.0 &  2.5 &  0.0 &  0.0 &  0.0 \\
Coronal                  &  6.9 &  4.3 &  3.0 &  2.3 &  2.1 & 10.9 \\
Classic Diagnostic       &  3.4 &  6.3 &  4.4 &  2.2 &  1.6 &  2.9 \\
Other Forbidden                & 12.6 & 12.9 & 16.9 & 11.8 &  8.6 & 12.5 \\
\multicolumn{7}{c}{\textit{Abundance}} \\
H $\textsc{i}$             & -- & -- & 16.8 & 17.2 & 16.1 & 28.8 \\
He $\textsc{i}$            & -- & -- &  3.2 &  3.6 &  8.4 &  5.4 \\
He $\textsc{ii}$           & -- & -- &  6.2 &  7.4 &  7.1 & 15.0 \\
Neutral C,N,O    & -- & -- &  2.1 &  0.0 &  0.0 &  0.0 \\
Coronal                  & -- & -- &  6.5 &  6.9 &  6.7 &  3.3 \\
Classic Diagnostic       & -- & -- &  4.8 &  1.9 &  2.7 &  2.8 \\
Other Forbidden                & -- & -- & 21.8 & 18.9 & 17.7 & 14.6 \\
\multicolumn{7}{c}{\textit{Luminosity}} \\
H $\textsc{i}$             & 24.9 & 15.3 & 10.2 & 22.9 & 23.0 & 16.5 \\
He $\textsc{i}$            & 14.5 & 16.3 &  9.7 &  8.8 & 11.9 &  8.2 \\
He $\textsc{ii}$           & 14.2 & 16.9 & 15.2 & 17.4 &  5.7 &  7.8 \\
Neutral C,N,O              &  0.0 &  0.0 &  2.6 &  0.0 &  0.0 &  0.0 \\
Coronal                  &  6.4 &  5.6 &  3.9 &  4.0 &  7.5 & 10.9 \\
Classic Diagnostic       &  3.1 &  6.0 &  4.6 &  4.1 &  5.1 &  2.9 \\
Other Forbidden                & 13.3 & 15.7 & 18.4 & 12.1 & 14.3 & 12.5 \\
\multicolumn{7}{c}{\textit{Temperature}} \\
H $\textsc{i}$             & 25.1 & 20.7 & 23.1 & 26.0 & 26.4 & 37.3 \\
He $\textsc{i}$            & 17.5 & 14.4 & 10.2 & 12.5 &  9.6 &  0.9 \\
He $\textsc{ii}$           & 18.1 & 18.9 &  1.9 & 18.4 & 21.3 & 27.7 \\
Neutral C,N,O              &  0.0 &  0.0 &  1.4 &  0.0 &  0.0 &  0.0 \\
Coronal                  &  5.0 &  5.0 &  5.8 &  3.0 &  2.6 &  3.7 \\
Classic Diagnostic       &  3.4 &  4.7 &  4.2 &  5.0 &  3.6 &  1.2 \\
Other Forbidden                & 11.3 & 13.7 & 19.5 & 10.8 &  9.9 &  7.7 \\
\enddata
}
\end{deluxetable}

\customref{tab:Diagnostic_Lines}{Tables \ref{tab:Diagnostic_Lines}} reveals a clear dominance of H and He lines in diagnosing Parameter-Space variables as  a direct result of the PCA methodology employed. As discussed in \customref{sec:Results-Eigenspectra & Correlations}{Section \ref{sec:Results-Eigenspectra & Correlations}}, H and He recombination lines often dominate the optical spectral features in nova shells and nebulae. Since PCA prioritizes global variance patterns within the dataset, H and He lines naturally dominate the mapping. Considering the emission process, there is a redundancy in the recombination transitions at low densities. However, the local emission roughly scales with the square of the density, being dominated by the highest density regions in the condensations and/or inner shell. In this case, self absorption is expected to be significant, and standard decrements are no longer followed. The intensities in all series are properly calculated and act as additional diagnostics inside the AI regressor. There is a very low or null contribution of weak neutral C,N,O permitted lines (except for a few cases at $320$ days). This is a consequence of our focus on evolved shells (when those transitions are usually fainter) and the brightness-driven selection method (see in \customref{sec:Methodology-Lines Fluxes and Variables Spaces}{Section \ref{sec:Methodology-Lines Fluxes and Variables Spaces}}), which does not impose prior knowledge about line formation but rather ranks the brightest ones.

\subsection{Machine Learning and Dimensionality Reduction Probe} \label{sec:Results-Machine Learning and Dimensionality Reduction Probe}

As a follow-up to our diagnostic line sets, we built a \customref{def:Random_Forest}{Random Forest Regressor} epoch-wise model for the post-eruption ages with complete Parameter-Space variable coverage. The goal is to constrain the physical parameters and abundances for each age using the information from the PCA-selected line sets. Other spectroscopic multivariate problems have been tackled with RFRs trained on the complete available data set \citep[e.g.][]{hong2025,qida2025}. In the present case, the limited sampling of parameters represents a computational limitation. On the other hand, a large number of observables (normalized line fluxes) are available for each nova age. 

The RFR prediction capability is evaluated by the mean absolute error (MAE), calculated from the differences between the parameter values found by the RFR (using the line fluxes only) and their actual model values, as set in the database spectrum synthesis. From now on, we will refer to the RFR trained with the Epoch-Selection 100 lines as the \textit{Goldstandard} (in terms of spectral information available) and compare its performance with the \customref{def:Top_n_Regressor}{Top $10$} PCA-based (refer to \customref{tab:top_lines_combined}{Table \ref{tab:top_lines_combined}}) regressor model. \customref{tab:mae_ranges}{Table \ref{tab:mae_ranges}} Summarize the MAE ranges for both regressors across all ages, clearly showing that the reduction in the number of lines used in the training from the $100$ brightest ones to the $10$ identified as most important by the PCA leads to slightly higher errors (especially for C,N,O abundance estimation) for most physical parameters; nevertheless, the metrics achieved are still lower than any other spectral regression method available, allowing us to conclude that the PCA line selection was successful, meaning a dimensionality reduction from $240$ to $10$ lines. Such an evaluation of regression performance with spectral information can only be conducted using a homogeneous synthetic database in which a large number of lines are available. Thanks to the previous PCA selection, it boils down to an observable line set.

Besides the excellent overall performance of both regressor classes mentioned above, higher uncertainties are found in estimating the condensed fraction, with systematically lower global variance explained by the model in comparison to the other physical variables. This limited performance does not prevent the regressor models from achieving a mean error of $30\%$ or less for the condensed mass fraction estimation, which seems to be a promising result for future improvement.

\begin{deluxetable}{lcccc}
\tablecaption{MAE Ranges: Goldstandard vs Top 10 Lines (All Ages)\label{tab:mae_ranges}}
\tablehead{
\colhead{Parameter} & 
\multicolumn{2}{c}{Goldstandard} & 
\multicolumn{2}{c}{Top 10 Lines} \\
\cline{2-3} \cline{4-5}
& \colhead{min} & \colhead{max} & \colhead{min} & \colhead{max}
}
\startdata
Mass (M$_\odot$)     & $1.03\times10^{-5}$ & $1.16\times10^{-5}$ & $1.20\times10^{-5}$ & $1.43\times10^{-5}$ \\
fc                   & 0.1357 & 0.1555 & 0.1575 & 0.1725 \\
Luminosity (dex) & 0.169 & 0.191 & 0.133 & 0.267 \\
Temperature (K)      & 6783.0 & 12997 & 10367 & 17436 \\
C,N,O abundance (dex)          & 0.0066 & 0.0177 & 0.0127 & 0.2038 \\
\enddata
\tablecomments{
MAE ranges across all models and all epochs (320-2560 days). The top 10 lines approach shows slightly higher errors for most parameters but maintains reasonable performance, demonstrating the efficacy of the diagnostic line selection.}
\end{deluxetable}

\begin{figure}
    \centering
    \includegraphics[width=1.0\textwidth]{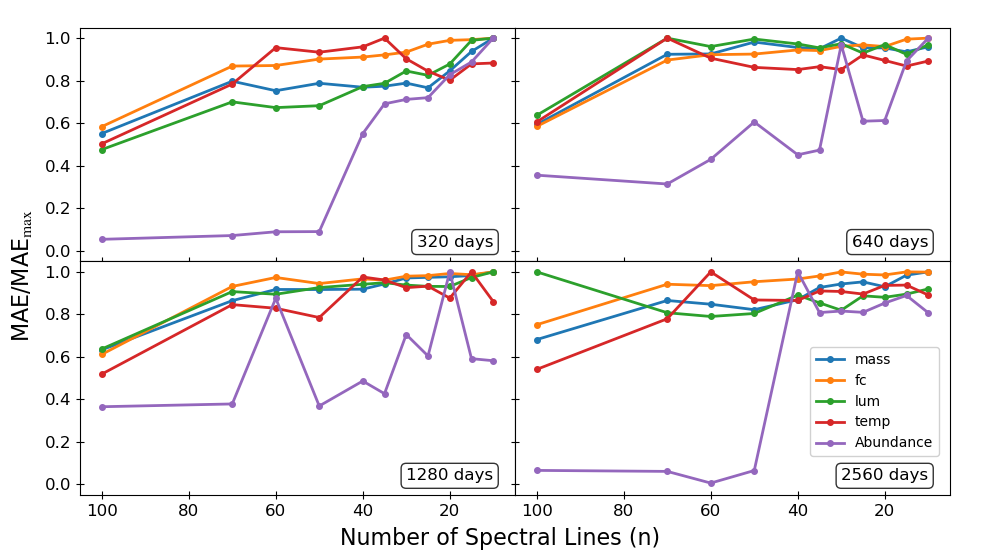}
    \caption{Normalized Mean Absolute Error achieved for the Goldstandard (100 lines) and Top $n$ RFR models}
    \label{fig:MAE_vs_n}
\end{figure}

For a more detailed view, \customref{fig:MAE_vs_n}{Figure \ref{fig:MAE_vs_n}} shows the epoch-wise MAE evolution with the decrease in the number of lines available for regressor training. It can be seen that, with the exception of the C,N,O abundance diagnosis, the multi-regressor dimensionality reduction, guided by the PCA, results in an evaluation metric degradation of no more than twice as large. Some grid variables even had increased local performance compared to larger $n$ regressors. The abundance retrieval shows some instability in the trend for some epochs,  but no more than a threefold increase is seen.

As a sanity check, we evaluated whether the bright Epoch-selection of 100 lines indeed does not limit the mapping capability of the AI models. For this purpose, for each post-eruption age, we built an RFR trained first using the entire nova Spectral-Space of 240 lines and compared its evaluation metrics with those of a regressor built with the respective 100 lines in Epoch-Selection. No significant difference was noted between the optimal models' performance trained with both line sets.

\subsection{Simulating Observations} \label{sec:Results-Simulating Observations}

As a final test of the framework's predictive power, we compute the median \customref{def:MAPE_and_MALE}{\textit{MAPE} and \textit{MALE}} for increasing levels of Gaussian noise added to the line intensities. These metrics quantify the expected predictive uncertainty when applying our regressor to observational data. Specifically, the \textit{MAPE} provides a percentage-based measure of accuracy for grid linear variables (namely T and fc), while the \textit{MALE} (expressed in dex) accounts for the order-of-magnitude variations inherent to grid logarithmic variables (M, L, and C,N,O abundance). By evaluating these metrics when using noisy line ratios, we test the robustness of the Top $10$ lines regressor model, demonstrating its ability to maintain reliable parameter retrieval.

\begin{figure}
    \centering
    \includegraphics[width=0.7\textwidth]{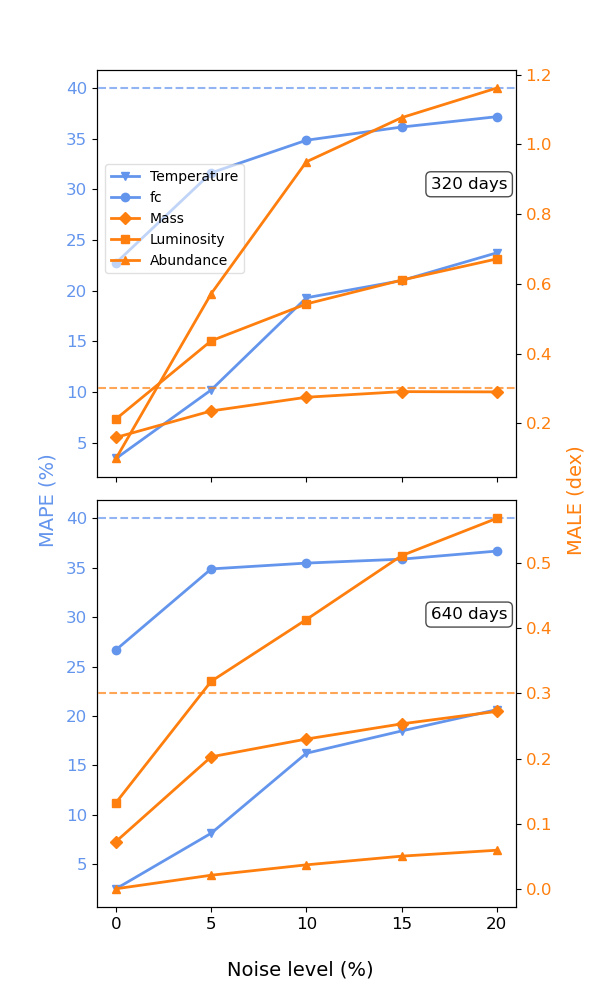}
    \caption{MAPE (left blue axis) and MALE (right orange axis) variables-wise with increasing simulated noise at $320$ (upper plot) and $640$ days (bottom plot). The dashed lines represents $40\%$ (blue) and $0.3 \ dex$ (orange) uncertainties.}
    \label{fig:Uncertainties_vs_Noise_320_&_640}
\end{figure}

\customref{fig:Uncertainties_vs_Noise_320_&_640}{Figure \ref{fig:Uncertainties_vs_Noise_320_&_640}} shows the Top $10$' performance metrics of the two earliest nebular phases of the NSBD; the behavior is the same for the two latest phases: even with a simulated $20\%$ flux error, a safe upper margin for exploration, the regressor is able to retrieve the central source temperature, shell condensed fraction, and ejected mass within $40\%$ and $0.3$ dex median uncertainties, which demonstrates the excellent regression power of this framework even with noisy data.

Concerning the luminosity and abundance regression, the first has the greatest uncertainty among the Parameter-Space, with an upper limit across post-eruption ages of $0.7$ dex. While this level of median error is somewhat greater than what some would consider acceptable, it is worth noting that this value is slightly above half of the grid's step, matching the expected level of type B uncertainty.

Finally, the abundance regression presents two distinct behaviors: at the two intermediate nebular ages, the regression was virtually perfect, achieving an uncertainty smaller than $0.06$ dex. On the other hand, at $320$ and $2560$ days, the uncertainty is enhanced to $1.16$ and $0.67$ dex, respectively. Despite these values not being far from the errors of current methods', the discrepancy in performance between these ages draws attention, showing that the method behaves differently depending on the shell physical conditions.

These results, along with the possibility of automating the framework constructed here, encourage us to improve the NSDB grid's resolution and add more parameters to better describe the white dwarf-shell system, such as an anisotropic radiation field component from an accretion disk and more detailed mass distribution abstractions. In addition to increasing the number of synthetic conditions per post-eruption age, the retrieval capability of the regressor may be further enhanced with additional nova observables, keeping in mind the application to targets from current and future spectroscopic synoptic surveys.

\section{Conclusions} \label{sec:Conclusions}

The Nova Synthetic Data Base, the first publicly available database of synthetic spectra for classical novae, is presented here. Generated with the CLOUDY based code \customref{def:RAINY3D}{RAINY3D}, this database contains $875+547$ nova remnant models with emission predicted from near-UV to the NIR \customref{Appendix:RAINY3D Spectral Lines Output}. The grids are constructed over a range of physical parameters that cover the most commonly reported values in the literature for classical novae in nebular phase. This kind of homogeneous grid, built with state-of-the-art photoionization models, is essential for the validation of robust multivariate and AI analysis methods, allowing for the development of multi-regressor machine learning algorithms with controlled physical data.

The framework's first step is to establish the eigenspectra for dimension reduction. The PCA analysis was carried out to construct a representative basis on which the spectroscopic variance can be related to the physical properties of the white dwarf-shell system mapped within the NSDB.

With a new maximum variance basis, the importance level of spectral lines for the retrieval of a grid's variable could be derived. The diagnostic significance rank is built by establishing a safe correlation threshold and the weight of a line in the description of eigenspectra; ultimately, this results in an efficient reduction of dimensionality from hundreds of predicted spectral lines to a restricted  \customref{tab:top_lines_combined}{observable set}, achievable only with the variance-focused PCA method.

With the most meaningful information derived from PCA, a Random Forest Regressor was built for each post-eruption age. The physical prediction capability is high, with the Mean Absolute Error (MAE) below grid sampling intervals for all derived parameters for the shell and central source. Comparing the regression metrics of the full line set and the most diagnostically capable ones, only a slight increase in error is observed, which still compares to or surpasses typical photoionization derived uncertainties.

The derived framework also proved robust,  demonstrating remarkable regression power in the presence of input flux data noise, with median errors below $0.3$ dex for the ejected mass, $0.7$ dex for luminosity, $40\%$ for the condensed mass fraction, and $25\%$ for the central source temperature, even with substantial $20\%$ flux uncertainty.

In light of these results, we propose the framework constructed here as a fully automatable spectral diagnostic method, suitable for large data sets available from current and future synoptic nova surveys. The extended proof of concept presented here also provides motivation to expand and improve the NSDB model grid.

\begin{acknowledgments}
This study was financed by CAPES process no. 8887.994579/2024-00 and by S\~ao Paulo Research Foundation (FAPESP) under grant 2019/08341-8. MPD thanks CNPq for support under grant \#305033. The authors acknowledge the National Laboratory for Scientific Computing (LNCC/MCTI,Brazil) for providing HPC resources of the SDumont supercomputer, which have contributed to the research results reported in this paper. The authors also thank Bruno Gerotti for his contribution to the calculation of NSDB. 

\end{acknowledgments}

\newpage
\appendix

\section{The Spectral-Space nova line selection} \label{Appendix:RAINY3D Spectral Lines Output}

This appendix provides the full list of the 241 spectral lines predicted by the RAINY3D photoionization models as output for the NSDB grids. The complete spectral-space selection, including ion identification and wavelengths, is presented in \customref{tab:Complete Spectral-Space selection of Nova lines}{Table \ref{tab:Complete Spectral-Space selection of Nova lines}}.

\begin{deluxetable*}{cccccccccc}
\tablecaption{Complete Spectral-Space selection of Nova lines.}
\label{tab:Complete Spectral-Space selection of Nova lines}
\tablehead{
\colhead{Element} & \colhead{$\lambda$ ($\lambda$\AA)} & 
\colhead{Element} & \colhead{$\lambda$ ($\lambda$\AA)} & 
\colhead{Element} & \colhead{$\lambda$ ($\lambda$\AA)} & 
\colhead{Element} & \colhead{$\lambda$ ($\lambda$\AA)} & 
\colhead{Element} & \colhead{$\lambda$ ($\lambda$\AA)}
}
\startdata
He $\textsc{ii}$ & $3203.04$ & Fe $\textsc{ii}$ & $3227.74$ & Fe $\textsc{ii}$ & $3255.89$ & Fe $\textsc{ii}$ & $3277.12$ & Fe $\textsc{ii}$ & $3281.29$ \\
$[$Ne $\textsc{iii}]$ & $3342.18$ & $[$Ne $\textsc{v}]$ & $3345.99$ & $[$Fe $\textsc{vi}]$ & $3813.55$ & $[$Ne $\textsc{v}]$ & $3426.03$ & O $\textsc{vi}$ & $3434.00$ \\
$[$Fe $\textsc{vi}]$ & $3492.10$ & $[$Fe $\textsc{vi}]$ & $3555.61$ & $[$Fe $\textsc{vii}]$ & $3586.32$ & $[$Fe $\textsc{vi}]$ & $3662.50$ & Ca $\textsc{ii}$ & $3706.02$ \\
$[$S $\textsc{iii}]$ & $3721.63$ & $[$O $\textsc{ii}]$ & $3726.03$ & H $\textsc{i}$ & $3734.36$ & Ca $\textsc{ii}$ & $3736.90$ & H $\textsc{i}$ & $3750.15$ \\
$[$Fe $\textsc{vii}]$ & $3758.92$ & H $\textsc{i}$ & $3770.63$ & H $\textsc{i}$ & $3797.89$ & H $\textsc{i}$ & $3835.38$ & He $\textsc{ii}$ & $3857.96$ \\
$[$Ne $\textsc{iii}]$ & $3868.76$ & He $\textsc{i}$ & $3888.63$ & Al $\textsc{ii}$ & $3900.67$ & He $\textsc{ii}$ & $3923.37$ & Ca $\textsc{ii}$ & $3933.66$ \\
$[$Ne $\textsc{iii}]$ & $3967.47$ & Ca $\textsc{ii}$ & $3968.47$ & H $\textsc{i}$ & $3970.07$ & $[$Fe $\textsc{xi}]$ & $3986.88$ & $[$Fe $\textsc{v}]$ & $4026.12$ \\
He $\textsc{i}$ & $4026.20$ & $[$S $\textsc{ii}]$ & $4068.60$ & $[$Fe $\textsc{v}]$ & $4071.24$ & $[$S $\textsc{ii}]$ & $4076.35$ & H $\textsc{i}$ & $4101.73$ \\
He $\textsc{i}$ & $4143.76$ & Fe $\textsc{ii}$ & $4178.96$ & $[$Fe $\textsc{v}]$ & $4180.59$ & C $\textsc{iii}$ & $4187.00$ & $[$Fe $\textsc{v}]$ & $4229.27$ \\
Fe $\textsc{ii}$ & $4233.17$ & $[$Fe $\textsc{ii}]$ & $4243.97$ & C $\textsc{ii}$ & $4267.00$ & $[$Fe $\textsc{ii}]$ & $4287.39$ & Fe $\textsc{ii}$ & $4305.89$ \\
H $\textsc{i}$ & $4340.46$ & Fe $\textsc{ii}$ & $4352.78$ & $[$Fe $\textsc{ii}]$ & $4359.33$ & $[$O $\textsc{iii}]$ & $4363.21$ & Fe $\textsc{ii}$ & $4384.32$ \\
He $\textsc{i}$ & $4387.93$ & Fe $\textsc{ii}$ & $4413.78$ & $[$Fe $\textsc{ii}]$ & $4416.27$ & He $\textsc{i}$ & $4437.55$ & $[$Fe $\textsc{ii}]$ & $4457.94$ \\
He $\textsc{i}$ & $4471.49$ & Mg $\textsc{ii}$ & $4481.15$ & Fe $\textsc{ii}$ & $4492.63$ & Fe $\textsc{ii}$ & $4514.90$ & He $\textsc{ii}$ & $4541.46$ \\
Fe $\textsc{ii}$ & $4555.89$ & Mg $\textsc{i}$ & $4571.10$ & Fe $\textsc{ii}$ & $4584.00$ & Fe $\textsc{ii}$ & $4629.34$ & N $\textsc{iii}$ & $4641.00$ \\
$[$Fe $\textsc{iii}]$ & $4658.01$ & C $\textsc{iv}$ & $4659.00$ & He $\textsc{ii}$ & $4685.64$ & $[$Fe $\textsc{iii}]$ & $4701.62$ & He $\textsc{i}$ & $4713.03$ \\
$[$Ar $\textsc{iv}]$ & $4740.12$ & $[$Fe $\textsc{iii}]$ & $4754.64$ & $[$Fe $\textsc{ii}]$ & $4814.53$ & H $\textsc{i}$ & $4861.33$ & $[$Fe $\textsc{iii}]$ & $4881.12$ \\
Fe $\textsc{vii}$ & $4893.37$ & He $\textsc{i}$ & $4921.93$ & Fe $\textsc{ii}$ & $4923.93$ & $[$Ca $\textsc{vii}]$ & $4939.56$ & $[$Fe $\textsc{vii}]$ & $4942.48$ \\
N $\textsc{v}$ & $4945.00$ & $[$O $\textsc{iii}]$ & $4958.91$ & $[$Fe $\textsc{vi}]$ & $4967.15$ & $[$Fe $\textsc{vi}]$ & $4971.72$ & $[$Fe $\textsc{vii}]$ & $4988.55$ \\
$[$O $\textsc{iii}]$ & $5006.84$ & He $\textsc{i}$ & $5015.68$ & Fe $\textsc{ii}$ & $5018.44$ & He $\textsc{i}$ & $5047.64$ & $[$Fe $\textsc{vi}]$ & $5145.76$ \\
$[$Fe $\textsc{ii}]$ & $5158.78$ & $[$Fe $\textsc{vii}]$ & $5158.89$ & Fe $\textsc{ii}$ & $5169.03$ & $[$Fe $\textsc{vi}]$ & $5176.05$ & $[$Fe $\textsc{ii}]$ & $5261.62$ \\
$[$Fe $\textsc{iii}]$ & $5270.40$ & Fe $\textsc{ii}$ & $5273.35$ & $[$Fe $\textsc{ii}]$ & $5276.00$ & $[$Fe $\textsc{vii}]$ & $5276.38$ & Fe $\textsc{ii}$ & $5284.11$ \\
O $\textsc{vi}$ & $5291.00$ & $[$Fe $\textsc{xiv}]$ & $5303.01$ & $[$Ca $\textsc{v}]$ & $5309.11$ & Fe $\textsc{ii}$ & $5316.62$ & Fe $\textsc{ii}$ & $5362.87$ \\
He $\textsc{ii}$ & $5411.37$ & Fe $\textsc{ii}$ & $5414.07$ & $[$Fe $\textsc{vi}]$ & $5424.23$ & Fe $\textsc{ii}$ & $5425.26$ & Fe $\textsc{ii}$ & $5527.34$ \\
$[$Ar $\textsc{x}]$ & $5534.02$ & Fe $\textsc{ii}$ & $5534.85$ & $[$O $\textsc{i}]$ & $5577.34$ & $[$Ca $\textsc{vii}]$ & $5618.76$ & $[$Fe $\textsc{vi}]$ & $5631.08$ \\
$[$Fe $\textsc{vi}]$ & $5676.96$ & N $\textsc{ii}$ & $5679.00$ & $[$Fe $\textsc{vii}]$ & $5720.71$ & N $\textsc{ii}$ & $5755.00$ & Ca $\textsc{vi}$ & $5765.38$ \\
He $\textsc{i}$ & $5875.64$ & Na $\textsc{i}$ & $5895.92$ & Fe $\textsc{ii}$ & $5991.38$ & Fe $\textsc{ii}$ & $6084.11$ & $[$Fe $\textsc{vii}]$ & $6086.97$ \\
$[$Ca $\textsc{v}]$ & $6086.37$ & He $\textsc{ii}$ & $6233.61$ & Fe $\textsc{ii}$ & $6248.06$ & $[$O $\textsc{i}]$ & $6300.30$ & He $\textsc{ii}$ & $6310.64$ \\
$[$S $\textsc{iii}]$ & $6312.06$ & Si $\textsc{ii}$ & $6347.11$ & $[$O $\textsc{i}]$ & $6363.78$ & Si $\textsc{ii}$ & $6371.37$ & $[$Fe $\textsc{x}]$ & $6374.54$ \\
$[$Ar $\textsc{v}]$ & $6435.12$ & $[$N $\textsc{ii}]$ & $6548.05$ & He $\textsc{ii}$ & $6559.91$ & H $\textsc{i}$ & $6562.81$ & $[$N $\textsc{ii}]$ & $6583.45$ \\
He $\textsc{i}$ & $6678.15$ & He $\textsc{ii}$ & $6682.98$ & Li $\textsc{i}$ & $6707.76$ & $[$S $\textsc{ii}]$ & $6716.44$ & $[$S $\textsc{ii}]$ & $6730.82$ \\
He $\textsc{ii}$ & $6890.67$ & $[$Ar $\textsc{v}]$ & $7005.83$ & He $\textsc{i}$ & $7065.22$ & $[$Ar $\textsc{iii}]$ & $7135.79$ & $[$Fe $\textsc{ii}]$ & $7155.16$ \\
$[$Ar $\textsc{iv}]$ & $7170.70$ & He $\textsc{ii}$ & $7177.28$ & Ca $\textsc{ii}$ & $7234.34$ & $[$Ar $\textsc{iv}]$ & $7237.77$ & $[$Ar $\textsc{iv}]$ & $7263.33$ \\
He $\textsc{i}$ & $7281.35$ & Ca $\textsc{ii}$ & $7291.47$ & Fe $\textsc{ii}$ & $7306.65$ & O $\textsc{ii}$ & $7329.67$ & $[$O $\textsc{ii}]$ & $7330.73$ \\
N $\textsc{i}$ & $7452.00$ & N $\textsc{iv}$ & $7582.00$ & He $\textsc{ii}$ & $7592.50$ & N $\textsc{iv}$ & $7703.00$ & O $\textsc{iv}$ & $7713.00$ \\
$[$S $\textsc{i}]$ & $7725.05$ & $[$Ar $\textsc{iii}]$ & $7751.11$ & O $\textsc{i}$ & $7773.00$ & $[$P $\textsc{ii}]$ & $7875.99$ & $[$Fe $\textsc{xi}]$ & $7891.87$ \\
Mg $\textsc{ii}$ & $7896.04$ & $[$Mn $\textsc{ix}]$ & $7968.49$ & N $\textsc{i}$ & $8212.00$ & He $\textsc{ii}$ & $8236.51$ & O $\textsc{i}$ & $8446.25$ \\
Ca $\textsc{ii}$ & $8498.02$ & H $\textsc{i}$ & $8502.44$ & Ca $\textsc{ii}$ & $8542.09$ & H $\textsc{i}$ & $8545.34$ & H $\textsc{i}$ & $8598.35$ \\
Ca $\textsc{ii}$ & $8662.14$ & H $\textsc{i}$ & $8664.98$ & N $\textsc{i}$ & $8692.00$ & $[$C $\textsc{i}]$ & $8727.13$ & H $\textsc{i}$ & $8750.43$ \\
Mg $\textsc{i}$ & $8806.76$ & H $\textsc{i}$ & $8862.74$ & H $\textsc{i}$ & $9014.87$ & N $\textsc{i}$ & $9028.92$ & N $\textsc{i}$ & $9060.48$ \\
$[$S $\textsc{iii}]$ & $9068.62$ & C $\textsc{i}$ & $9088.00$ & Mg $\textsc{ii}$ & $9218.25$ & H $\textsc{i}$ & $9228.97$ & Mg $\textsc{ii}$ & $9244.27$ \\
O $\textsc{i}$ & $9264.00$ & He $\textsc{ii}$ & $9344.62$ & N $\textsc{i}$ & $9395.85$ & $[$S $\textsc{iii}]$ & $9530.62$ & He $\textsc{ii}$ & $9541.71$ \\
H $\textsc{i}$ & $9545.93$ & C $\textsc{i}$ & $9658.00$ & He $\textsc{ii}$ & $9761.80$ & He $\textsc{ii}$ & $10044.90$ & H $\textsc{i}$ & $10049.30$ \\
N $\textsc{i}$ & $10117.00$ & He $\textsc{ii}$ & $10123.30$ & N $\textsc{i}$ & $10401.30$ & N $\textsc{i}$ & $10525.00$ & C $\textsc{i}$ & $10695.00$ \\
He $\textsc{i}$ & $10830.30$ & H $\textsc{i}$ & $10938.00$ & O $\textsc{i}$ & $11286.30$ & H $\textsc{i}$ & $12818.00$ & C $\textsc{i}$ & $14542.50$ \\
H $\textsc{i}$ & $15438.80$ & H $\textsc{i}$ & $15556.30$ & H $\textsc{i}$ & $15700.60$ & H $\textsc{i}$ & $15880.40$ & H $\textsc{i}$ & $16109.20$ \\
H $\textsc{i}$ & $16407.10$ & H $\textsc{i}$ & $16806.40$ & He $\textsc{i}$ & $17002.70$ & H $\textsc{i}$ & $17362.00$ & H $\textsc{i}$ & $18174.00$ \\
H $\textsc{i}$ & $18751.00$ & H $\textsc{i}$ & $19445.40$ & He $\textsc{ii}$ & $20372.50$ & $[$Al $\textsc{ix}]$ & $20444.40$ & He $\textsc{i}$ & $20581.30$ \\
He $\textsc{i}$ & $21130.30$ & H $\textsc{i}$ & $21655.10$ & He $\textsc{ii}$ & $21884.30$ & $[$Ca $\textsc{viii}]$ & $23211.70$ & He $\textsc{ii}$ & $23463.10$ \\
$[$Si $\textsc{vii}]$ & $24807.10$ &  &  &  &  &  &  &  &  \\
\enddata
\end{deluxetable*}

\section{Brightest 100 line selections per age} \label{Appendix:Epoch-Selections}

Epoch-Selections in increasing order of wavelength are shown in \customref{tab:Epoch-Selection_80d}{Tables \ref{tab:Epoch-Selection_80d}}-\customref{tab:Epoch-Selection_2560d}{\ref{tab:Epoch-Selection_2560d}}.

\begin{deluxetable*}{cccccccccc}
\tablecaption{Epoch-Selection for 80 days age.}
\label{tab:Epoch-Selection_80d}
\tablehead{
\colhead{Element} & \colhead{$\lambda$ ($\lambda$\AA)} & 
\colhead{Element} & \colhead{$\lambda$ ($\lambda$\AA)} & 
\colhead{Element} & \colhead{$\lambda$ ($\lambda$\AA)} & 
\colhead{Element} & \colhead{$\lambda$ ($\lambda$\AA)} & 
\colhead{Element} & \colhead{$\lambda$ ($\lambda$\AA)}
}
\startdata
He $\textsc{ii}$ & $3203.04$ & $[$Ne $\textsc{iii}]$ & $3342.18$ & $[$Ne $\textsc{v}]$ & $3345.99$ & $[$Ne $\textsc{v}]$ & $3426.03$ & O $\textsc{vi}$ & $3434.00$ \\
$[$Fe $\textsc{vi}]$ & $3492.10$ & $[$Fe $\textsc{vi}]$ & $3555.61$ & $[$Fe $\textsc{vii}]$ & $3586.32$ & $[$Fe $\textsc{vi}]$ & $3662.50$ & H $\textsc{i}$ & $3734.36$ \\
H $\textsc{i}$ & $3750.15$ & $[$Fe $\textsc{vii}]$ & $3758.92$ & H $\textsc{i}$ & $3770.63$ & H $\textsc{i}$ & $3797.89$ & $[$Fe $\textsc{vi}]$ & $3813.55$ \\
H $\textsc{i}$ & $3835.38$ & $[$Ne $\textsc{iii}]$ & $3868.76$ & He $\textsc{i}$ & $3888.63$ & He $\textsc{ii}$ & $3923.37$ & $[$Ne $\textsc{iii}]$ & $3967.47$ \\
H $\textsc{i}$ & $3970.07$ & $[$Fe $\textsc{xi}]$ & $3986.88$ & He $\textsc{i}$ & $4026.20$ & $[$Fe $\textsc{v}]$ & $4071.24$ & H $\textsc{i}$ & $4101.73$ \\
$[$Fe $\textsc{v}]$ & $4180.59$ & H $\textsc{i}$ & $4340.46$ & $[$O $\textsc{iii}]$ & $4363.21$ & He $\textsc{i}$ & $4387.93$ & He $\textsc{i}$ & $4471.49$ \\
He $\textsc{ii}$ & $4541.46$ & C $\textsc{iv}$ & $4659.00$ & He $\textsc{ii}$ & $4685.64$ & He $\textsc{i}$ & $4713.03$ & $[$Ar $\textsc{iv}]$ & $4740.12$ \\
Fe $\textsc{vii}$ & $4893.37$ & He $\textsc{i}$ & $4921.93$ & $[$Ca $\textsc{vii}]$ & $4939.56$ & $[$Fe $\textsc{vii}]$ & $4942.48$ & N $\textsc{v}$ & $4945.00$ \\
$[$O $\textsc{iii}]$ & $4958.91$ & $[$Fe $\textsc{vi}]$ & $4967.15$ & $[$Fe $\textsc{vii}]$ & $4988.55$ & $[$O $\textsc{iii}]$ & $5006.84$ & He $\textsc{i}$ & $5015.68$ \\
$[$Fe $\textsc{vi}]$ & $5145.76$ & $[$Fe $\textsc{vii}]$ & $5158.89$ & $[$Fe $\textsc{vi}]$ & $5176.05$ & $[$Fe $\textsc{vii}]$ & $5276.38$ & O $\textsc{vi}$ & $5291.00$ \\
$[$Ca $\textsc{v}]$ & $5309.11$ & He $\textsc{ii}$ & $5411.37$ & $[$Fe $\textsc{vi}]$ & $5424.23$ & $[$Ar $\textsc{x}]$ & $5534.02$ & $[$Ca $\textsc{vii}]$ & $5618.76$ \\
$[$Fe $\textsc{vi}]$ & $5631.08$ & $[$Fe $\textsc{vi}]$ & $5676.96$ & $[$Fe $\textsc{vii}]$ & $5720.71$ & $[$N $\textsc{ii}]$ & $5755.00$ & He $\textsc{i}$ & $5875.64$ \\
$[$Ca $\textsc{v}]$ & $6086.37$ & $[$Fe $\textsc{vii}]$ & $6086.97$ & $[$S $\textsc{iii}]$ & $6312.06$ & $[$Fe $\textsc{x}]$ & $6374.54$ & $[$Ar $\textsc{v}]$ & $6435.12$ \\
He $\textsc{ii}$ & $6559.91$ & H $\textsc{i}$ & $6562.81$ & He $\textsc{i}$ & $6678.15$ & He $\textsc{ii}$ & $6682.98$ & He $\textsc{ii}$ & $6890.67$ \\
$[$Ar $\textsc{v}]$ & $7005.83$ & He $\textsc{i}$ & $7065.22$ & $[$Ar $\textsc{iii}]$ & $7135.79$ & $[$Ar $\textsc{iv}]$ & $7170.70$ & He $\textsc{ii}$ & $7177.28$ \\
He $\textsc{i}$ & $7281.35$ & He $\textsc{ii}$ & $7592.50$ & $[$Fe $\textsc{xi}]$ & $7891.87$ & $[$Mn $\textsc{ix}]$ & $7968.49$ & He $\textsc{ii}$ & $8236.51$ \\
H $\textsc{i}$ & $8750.43$ & H $\textsc{i}$ & $8862.74$ & H $\textsc{i}$ & $9014.87$ & H $\textsc{i}$ & $9228.97$ & He $\textsc{ii}$ & $9344.62$ \\
$[$S $\textsc{iii}]$ & $9530.62$ & H $\textsc{i}$ & $9545.93$ & H $\textsc{i}$ & $10049.30$ & He $\textsc{ii}$ & $10123.30$ & He $\textsc{i}$ & $10830.30$ \\
H $\textsc{i}$ & $10938.00$ & H $\textsc{i}$ & $12818.00$ & H $\textsc{i}$ & $18174.00$ & H $\textsc{i}$ & $18751.00$ & H $\textsc{i}$ & $19445.40$ \\
$[$Al $\textsc{ix}]$ & $20444.40$ & He $\textsc{i}$ & $20581.30$ & H $\textsc{i}$ & $21655.10$ & $[$Ca $\textsc{viii}]$ & $23211.70$ & $[$Si $\textsc{vii}]$ & $24807.10$ \\
\enddata
\end{deluxetable*}

\begin{deluxetable*}{cccccccccc}
\tablecaption{Epoch-Selection for 160 days age.}
\tablehead{
\colhead{Element} & \colhead{$\lambda$ ($\lambda$\AA)} & 
\colhead{Element} & \colhead{$\lambda$ ($\lambda$\AA)} & 
\colhead{Element} & \colhead{$\lambda$ ($\lambda$\AA)} & 
\colhead{Element} & \colhead{$\lambda$ ($\lambda$\AA)} & 
\colhead{Element} & \colhead{$\lambda$ ($\lambda$\AA)}
}
\startdata
He $\textsc{ii}$ & $3203.04$ & $[$Ne $\textsc{iii}]$ & $3342.18$ & $[$Ne $\textsc{v}]$ & $3345.99$ & $[$Ne $\textsc{v}]$ & $3426.03$ & O $\textsc{vi}$ & $3434.00$ \\
$[$Fe $\textsc{vi}]$ & $3492.10$ & $[$Fe $\textsc{vi}]$ & $3555.61$ & $[$Fe $\textsc{vii}]$ & $3586.32$ & $[$Fe $\textsc{vi}]$ & $3662.50$ & $[$S $\textsc{iii}]$ & $3721.63$ \\
H $\textsc{i}$ & $3734.36$ & H $\textsc{i}$ & $3750.15$ & $[$Fe $\textsc{vii}]$ & $3758.92$ & H $\textsc{i}$ & $3770.63$ & H $\textsc{i}$ & $3797.89$ \\
$[$Fe $\textsc{vi}]$ & $3813.55$ & H $\textsc{i}$ & $3835.38$ & He $\textsc{ii}$ & $3857.96$ & $[$Ne $\textsc{iii}]$ & $3868.76$ & He $\textsc{i}$ & $3888.63$ \\
He $\textsc{ii}$ & $3923.37$ & $[$Ne $\textsc{iii}]$ & $3967.47$ & H $\textsc{i}$ & $3970.07$ & $[$Fe $\textsc{xi}]$ & $3986.88$ & He $\textsc{i}$ & $4026.20$ \\
$[$Fe $\textsc{v}]$ & $4071.24$ & H $\textsc{i}$ & $4101.73$ & $[$Fe $\textsc{v}]$ & $4180.59$ & H $\textsc{i}$ & $4340.46$ & $[$O $\textsc{iii}]$ & $4363.21$ \\
He $\textsc{i}$ & $4471.49$ & He $\textsc{ii}$ & $4541.46$ & C $\textsc{iv}$ & $4659.00$ & He $\textsc{ii}$ & $4685.64$ & He $\textsc{i}$ & $4713.03$ \\
$[$Ar $\textsc{iv}]$ & $4740.12$ & Fe $\textsc{vii}$ & $4893.37$ & He $\textsc{i}$ & $4921.93$ & $[$Ca $\textsc{vii}]$ & $4939.56$ & $[$Fe $\textsc{vii}]$ & $4942.48$ \\
N $\textsc{v}$ & $4945.00$ & $[$O $\textsc{iii}]$ & $4958.91$ & $[$Fe $\textsc{vi}]$ & $4967.15$ & $[$Fe $\textsc{vii}]$ & $4988.55$ & $[$O $\textsc{iii}]$ & $5006.84$ \\
He $\textsc{i}$ & $5015.68$ & $[$Fe $\textsc{vi}]$ & $5145.76$ & $[$Fe $\textsc{vii}]$ & $5158.89$ & $[$Fe $\textsc{vi}]$ & $5176.05$ & $[$Fe $\textsc{vii}]$ & $5276.38$ \\
O $\textsc{vi}$ & $5291.00$ & $[$Fe $\textsc{xiv}]$ & $5303.01$ & $[$Ca $\textsc{v}]$ & $5309.11$ & He $\textsc{ii}$ & $5411.37$ & $[$Fe $\textsc{vi}]$ & $5424.23$ \\
$[$Ar $\textsc{x}]$ & $5534.02$ & $[$Ca $\textsc{vii}]$ & $5618.76$ & $[$Fe $\textsc{vi}]$ & $5631.08$ & $[$Fe $\textsc{vi}]$ & $5676.96$ & $[$Fe $\textsc{vii}]$ & $5720.71$ \\
$[$N $\textsc{ii}]$ & $5755.00$ & He $\textsc{i}$ & $5875.64$ & $[$Ca $\textsc{v}]$ & $6086.37$ & $[$Fe $\textsc{vii}]$ & $6086.97$ & $[$S $\textsc{iii}]$ & $6312.06$ \\
$[$Fe $\textsc{x}]$ & $6374.54$ & $[$Ar $\textsc{v}]$ & $6435.12$ & $[$N $\textsc{ii}]$ & $6548.05$ & He $\textsc{ii}$ & $6559.91$ & H $\textsc{i}$ & $6562.81$ \\
$[$N $\textsc{ii}]$ & $6583.45$ & He $\textsc{i}$ & $6678.15$ & He $\textsc{ii}$ & $6682.98$ & He $\textsc{ii}$ & $6890.67$ & $[$Ar $\textsc{v}]$ & $7005.83$ \\
He $\textsc{i}$ & $7065.22$ & $[$Ar $\textsc{iii}]$ & $7135.79$ & He $\textsc{ii}$ & $7177.28$ & He $\textsc{i}$ & $7281.35$ & He $\textsc{ii}$ & $7592.50$ \\
$[$Fe $\textsc{xi}]$ & $7891.87$ & $[$Mn $\textsc{ix}]$ & $7968.49$ & He $\textsc{ii}$ & $8236.51$ & H $\textsc{i}$ & $9014.87$ & $[$S $\textsc{iii}]$ & $9068.62$ \\
H $\textsc{i}$ & $9228.97$ & He $\textsc{ii}$ & $9344.62$ & $[$S $\textsc{iii}]$ & $9530.62$ & H $\textsc{i}$ & $9545.93$ & H $\textsc{i}$ & $10049.30$ \\
He $\textsc{ii}$ & $10123.30$ & He $\textsc{i}$ & $10830.30$ & H $\textsc{i}$ & $10938.00$ & H $\textsc{i}$ & $12818.00$ & H $\textsc{i}$ & $18751.00$ \\
$[$Al $\textsc{ix}]$ & $20444.40$ & He $\textsc{i}$ & $20581.30$ & H $\textsc{i}$ & $21655.10$ & $[$Ca $\textsc{viii}]$ & $23211.70$ & $[$Si $\textsc{vii}]$ & $24807.10$ \\
\enddata
\end{deluxetable*}

\begin{deluxetable*}{cccccccccc}
\tablecaption{Epoch-Selection for 320 days age.}
\tablehead{
\colhead{Element} & \colhead{$\lambda$ ($\lambda$\AA)} & 
\colhead{Element} & \colhead{$\lambda$ ($\lambda$\AA)} & 
\colhead{Element} & \colhead{$\lambda$ ($\lambda$\AA)} & 
\colhead{Element} & \colhead{$\lambda$ ($\lambda$\AA)} & 
\colhead{Element} & \colhead{$\lambda$ ($\lambda$\AA)}
}
\startdata
He $\textsc{ii}$ & $3203.04$ & $[$Ne $\textsc{v}]$ & $3345.99$ & $[$Ne $\textsc{v}]$ & $3426.03$ & O $\textsc{vi}$ & $3434.00$ & $[$Fe $\textsc{vi}]$ & $3492.10$ \\
$[$Fe $\textsc{vi}]$ & $3555.61$ & $[$Fe $\textsc{vii}]$ & $3586.32$ & $[$Fe $\textsc{vi}]$ & $3662.50$ & $[$O $\textsc{ii}]$ & $3726.03$ & H $\textsc{i}$ & $3734.36$ \\
H $\textsc{i}$ & $3750.15$ & $[$Fe $\textsc{vii}]$ & $3758.92$ & H $\textsc{i}$ & $3770.63$ & H $\textsc{i}$ & $3797.89$ & $[$Fe $\textsc{vi}]$ & $3813.55$ \\
H $\textsc{i}$ & $3835.38$ & $[$Ne $\textsc{iii}]$ & $3868.76$ & He $\textsc{i}$ & $3888.63$ & $[$Ne $\textsc{iii}]$ & $3967.47$ & H $\textsc{i}$ & $3970.07$ \\
$[$Fe $\textsc{xi}]$ & $3986.88$ & He $\textsc{i}$ & $4026.20$ & $[$Fe $\textsc{v}]$ & $4071.24$ & H $\textsc{i}$ & $4101.73$ & $[$Fe $\textsc{v}]$ & $4180.59$ \\
C $\textsc{iii}$ & $4187.00$ & C $\textsc{ii}$ & $4267.00$ & H $\textsc{i}$ & $4340.46$ & $[$O $\textsc{iii}]$ & $4363.21$ & He $\textsc{i}$ & $4471.49$ \\
He $\textsc{ii}$ & $4541.46$ & N $\textsc{iii}$ & $4641.00$ & C $\textsc{iv}$ & $4659.00$ & He $\textsc{ii}$ & $4685.64$ & $[$Ar $\textsc{iv}]$ & $4740.12$ \\
Fe $\textsc{vii}$ & $4893.37$ & $[$Ca $\textsc{vii}]$ & $4939.56$ & $[$Fe $\textsc{vii}]$ & $4942.48$ & N $\textsc{v}$ & $4945.00$ & $[$O $\textsc{iii}]$ & $4958.91$ \\
$[$Fe $\textsc{vi}]$ & $4967.15$ & $[$Fe $\textsc{vii}]$ & $4988.55$ & $[$O $\textsc{iii}]$ & $5006.84$ & He $\textsc{i}$ & $5015.68$ & $[$Fe $\textsc{vi}]$ & $5145.76$ \\
$[$Fe $\textsc{vii}]$ & $5158.89$ & $[$Fe $\textsc{vi}]$ & $5176.05$ & $[$Fe $\textsc{vii}]$ & $5276.38$ & O $\textsc{vi}$ & $5291.00$ & $[$Fe $\textsc{xiv}]$ & $5303.01$ \\
$[$Ca $\textsc{v}]$ & $5309.11$ & He $\textsc{ii}$ & $5411.37$ & $[$Fe $\textsc{vi}]$ & $5424.23$ & $[$Ar $\textsc{x}]$ & $5534.02$ & $[$Ca $\textsc{vii}]$ & $5618.76$ \\
$[$Fe $\textsc{vi}]$ & $5631.08$ & $[$Fe $\textsc{vi}]$ & $5676.96$ & N $\textsc{ii}$ & $5679.00$ & $[$Fe $\textsc{vii}]$ & $5720.71$ & $[$N $\textsc{ii}]$ & $5755.00$ \\
He $\textsc{i}$ & $5875.64$ & $[$Ca $\textsc{v}]$ & $6086.37$ & $[$Fe $\textsc{vii}]$ & $6086.97$ & $[$S $\textsc{iii}]$ & $6312.06$ & $[$Fe $\textsc{x}]$ & $6374.54$ \\
$[$Ar $\textsc{v}]$ & $6435.12$ & $[$N $\textsc{ii}]$ & $6548.05$ & He $\textsc{ii}$ & $6559.91$ & H $\textsc{i}$ & $6562.81$ & $[$N $\textsc{ii}]$ & $6583.45$ \\
He $\textsc{i}$ & $6678.15$ & $[$Ar $\textsc{v}]$ & $7005.83$ & He $\textsc{i}$ & $7065.22$ & $[$Ar $\textsc{iii}]$ & $7135.79$ & He $\textsc{ii}$ & $7177.28$ \\
O $\textsc{ii}$ & $7329.67$ & $[$O $\textsc{ii}]$ & $7330.73$ & N $\textsc{iv}$ & $7582.00$ & He $\textsc{ii}$ & $7592.50$ & N $\textsc{iv}$ & $7703.00$ \\
O $\textsc{iv}$ & $7713.00$ & O $\textsc{i}$ & $7773.00$ & $[$Fe $\textsc{xi}]$ & $7891.87$ & $[$Mn $\textsc{ix}]$ & $7968.49$ & He $\textsc{ii}$ & $8236.51$ \\
N $\textsc{i}$ & $8692.00$ & $[$S $\textsc{iii}]$ & $9068.62$ & O $\textsc{i}$ & $9264.00$ & He $\textsc{ii}$ & $9344.62$ & $[$S $\textsc{iii}]$ & $9530.62$ \\
H $\textsc{i}$ & $9545.93$ & H $\textsc{i}$ & $10049.30$ & He $\textsc{ii}$ & $10123.30$ & He $\textsc{i}$ & $10830.30$ & H $\textsc{i}$ & $10938.00$ \\
H $\textsc{i}$ & $12818.00$ & H $\textsc{i}$ & $18751.00$ & $[$Al $\textsc{ix}]$ & $20444.40$ & $[$Ca $\textsc{viii}]$ & $23211.70$ & $[$Si $\textsc{vii}]$ & $24807.10$ \\
\enddata
\end{deluxetable*}

\begin{deluxetable*}{cccccccccc}
\tablecaption{Epoch-Selection for 640 days age.}
\tablehead{
\colhead{Element} & \colhead{$\lambda$ ($\lambda$\AA)} & 
\colhead{Element} & \colhead{$\lambda$ ($\lambda$\AA)} & 
\colhead{Element} & \colhead{$\lambda$ ($\lambda$\AA)} & 
\colhead{Element} & \colhead{$\lambda$ ($\lambda$\AA)} & 
\colhead{Element} & \colhead{$\lambda$ ($\lambda$\AA)}
}
\startdata
He $\textsc{ii}$ & $3203.04$ & $[$Ne $\textsc{v}]$ & $3345.99$ & $[$Ne $\textsc{v}]$ & $3426.03$ & O $\textsc{vi}$ & $3434.00$ & $[$Fe $\textsc{vi}]$ & $3492.10$ \\
$[$Fe $\textsc{vi}]$ & $3555.61$ & $[$Fe $\textsc{vii}]$ & $3586.32$ & $[$Fe $\textsc{vi}]$ & $3662.50$ & H $\textsc{i}$ & $3734.36$ & H $\textsc{i}$ & $3750.15$ \\
$[$Fe $\textsc{vii}]$ & $3758.92$ & H $\textsc{i}$ & $3770.63$ & H $\textsc{i}$ & $3797.89$ & $[$Fe $\textsc{vi}]$ & $3813.55$ & H $\textsc{i}$ & $3835.38$ \\
He $\textsc{ii}$ & $3857.96$ & $[$Ne $\textsc{iii}]$ & $3868.76$ & He $\textsc{i}$ & $3888.63$ & He $\textsc{ii}$ & $3923.37$ & $[$Ne $\textsc{iii}]$ & $3967.47$ \\
H $\textsc{i}$ & $3970.07$ & $[$Fe $\textsc{xi}]$ & $3986.88$ & He $\textsc{i}$ & $4026.20$ & $[$Fe $\textsc{v}]$ & $4071.24$ & H $\textsc{i}$ & $4101.73$ \\
$[$Fe $\textsc{v}]$ & $4180.59$ & C $\textsc{iii}$ & $4187.00$ & C $\textsc{ii}$ & $4267.00$ & H $\textsc{i}$ & $4340.46$ & $[$O $\textsc{iii}]$ & $4363.21$ \\
He $\textsc{i}$ & $4471.49$ & He $\textsc{ii}$ & $4541.46$ & N $\textsc{iii}$ & $4641.00$ & C $\textsc{iv}$ & $4659.00$ & He $\textsc{ii}$ & $4685.64$ \\
$[$Ar $\textsc{iv}]$ & $4740.12$ & Fe $\textsc{vii}$ & $4893.37$ & $[$Ca $\textsc{vii}]$ & $4939.56$ & $[$Fe $\textsc{vii}]$ & $4942.48$ & N $\textsc{v}$ & $4945.00$ \\
$[$O $\textsc{iii}]$ & $4958.91$ & $[$Fe $\textsc{vi}]$ & $4967.15$ & $[$Fe $\textsc{vii}]$ & $4988.55$ & $[$O $\textsc{iii}]$ & $5006.84$ & He $\textsc{i}$ & $5015.68$ \\
$[$Fe $\textsc{vi}]$ & $5145.76$ & $[$Fe $\textsc{vii}]$ & $5158.89$ & $[$Fe $\textsc{vi}]$ & $5176.05$ & $[$Fe $\textsc{vii}]$ & $5276.38$ & O $\textsc{vi}$ & $5291.00$ \\
$[$Fe $\textsc{xiv}]$ & $5303.01$ & $[$Ca $\textsc{v}]$ & $5309.11$ & He $\textsc{ii}$ & $5411.37$ & $[$Fe $\textsc{vi}]$ & $5424.23$ & $[$Ar $\textsc{x}]$ & $5534.02$ \\
$[$Ca $\textsc{vii}]$ & $5618.76$ & $[$Fe $\textsc{vi}]$ & $5631.08$ & $[$Fe $\textsc{vi}]$ & $5676.96$ & N $\textsc{ii}$ & $5679.00$ & $[$Fe $\textsc{vii}]$ & $5720.71$ \\
He $\textsc{i}$ & $5875.64$ & $[$Ca $\textsc{v}]$ & $6086.37$ & $[$Fe $\textsc{vii}]$ & $6086.97$ & $[$Fe $\textsc{x}]$ & $6374.54$ & $[$Ar $\textsc{v}]$ & $6435.12$ \\
$[$N $\textsc{ii}]$ & $6548.05$ & He $\textsc{ii}$ & $6559.91$ & H $\textsc{i}$ & $6562.81$ & $[$N $\textsc{ii}]$ & $6583.45$ & He $\textsc{i}$ & $6678.15$ \\
He $\textsc{ii}$ & $6682.98$ & He $\textsc{ii}$ & $6890.67$ & $[$Ar $\textsc{v}]$ & $7005.83$ & He $\textsc{i}$ & $7065.22$ & He $\textsc{ii}$ & $7177.28$ \\
N $\textsc{iv}$ & $7582.00$ & He $\textsc{ii}$ & $7592.50$ & N $\textsc{iv}$ & $7703.00$ & O $\textsc{iv}$ & $7713.00$ & $[$Fe $\textsc{xi}]$ & $7891.87$ \\
$[$Mn $\textsc{ix}]$ & $7968.49$ & He $\textsc{ii}$ & $8236.51$ & H $\textsc{i}$ & $8862.74$ & H $\textsc{i}$ & $9014.87$ & $[$S $\textsc{iii}]$ & $9068.62$ \\
H $\textsc{i}$ & $9228.97$ & He $\textsc{ii}$ & $9344.62$ & $[$S $\textsc{iii}]$ & $9530.62$ & H $\textsc{i}$ & $9545.93$ & H $\textsc{i}$ & $10049.30$ \\
He $\textsc{ii}$ & $10123.30$ & He $\textsc{i}$ & $10830.30$ & H $\textsc{i}$ & $10938.00$ & H $\textsc{i}$ & $12818.00$ & H $\textsc{i}$ & $18751.00$ \\
H $\textsc{i}$ & $19445.40$ & $[$Al $\textsc{ix}]$ & $20444.40$ & H $\textsc{i}$ & $21655.10$ & $[$Ca $\textsc{viii}]$ & $23211.70$ & $[$Si $\textsc{vii}]$ & $24807.10$ \\
\enddata
\end{deluxetable*}

\begin{deluxetable*}{cccccccccc}
\tablecaption{Epoch-Selection for 1280 days age.}
\tablehead{
\colhead{Element} & \colhead{$\lambda$ ($\lambda$\AA)} & 
\colhead{Element} & \colhead{$\lambda$ ($\lambda$\AA)} & 
\colhead{Element} & \colhead{$\lambda$ ($\lambda$\AA)} & 
\colhead{Element} & \colhead{$\lambda$ ($\lambda$\AA)} & 
\colhead{Element} & \colhead{$\lambda$ ($\lambda$\AA)}
}
\startdata
He $\textsc{ii}$ & $3203.04$ & $[$Ne $\textsc{v}]$ & $3345.99$ & $[$Ne $\textsc{v}]$ & $3426.03$ & O $\textsc{vi}$ & $3434.00$ & $[$Fe $\textsc{vi}]$ & $3492.10$ \\
$[$Fe $\textsc{vi}]$ & $3555.61$ & $[$Fe $\textsc{vii}]$ & $3586.32$ & $[$Fe $\textsc{vi}]$ & $3662.50$ & H $\textsc{i}$ & $3734.36$ & H $\textsc{i}$ & $3750.15$ \\
$[$Fe $\textsc{vii}]$ & $3758.92$ & H $\textsc{i}$ & $3770.63$ & H $\textsc{i}$ & $3797.89$ & $[$Fe $\textsc{vi}]$ & $3813.55$ & H $\textsc{i}$ & $3835.38$ \\
He $\textsc{ii}$ & $3857.96$ & $[$Ne $\textsc{iii}]$ & $3868.76$ & He $\textsc{i}$ & $3888.63$ & He $\textsc{ii}$ & $3923.37$ & $[$Ne $\textsc{iii}]$ & $3967.47$ \\
H $\textsc{i}$ & $3970.07$ & $[$Fe $\textsc{xi}]$ & $3986.88$ & He $\textsc{i}$ & $4026.20$ & $[$Fe $\textsc{v}]$ & $4071.24$ & H $\textsc{i}$ & $4101.73$ \\
$[$Fe $\textsc{v}]$ & $4180.59$ & C $\textsc{iii}$ & $4187.00$ & C $\textsc{ii}$ & $4267.00$ & H $\textsc{i}$ & $4340.46$ & $[$O $\textsc{iii}]$ & $4363.21$ \\
He $\textsc{i}$ & $4471.49$ & He $\textsc{ii}$ & $4541.46$ & N $\textsc{iii}$ & $4641.00$ & C $\textsc{iv}$ & $4659.00$ & He $\textsc{ii}$ & $4685.64$ \\
$[$Ar $\textsc{iv}]$ & $4740.12$ & Fe $\textsc{vii}$ & $4893.37$ & He $\textsc{i}$ & $4921.93$ & $[$Ca $\textsc{vii}]$ & $4939.56$ & $[$Fe $\textsc{vii}]$ & $4942.48$ \\
N $\textsc{v}$ & $4945.00$ & $[$O $\textsc{iii}]$ & $4958.91$ & $[$Fe $\textsc{vi}]$ & $4967.15$ & $[$Fe $\textsc{vii}]$ & $4988.55$ & $[$O $\textsc{iii}]$ & $5006.84$ \\
$[$Fe $\textsc{vi}]$ & $5145.76$ & $[$Fe $\textsc{vii}]$ & $5158.89$ & $[$Fe $\textsc{vi}]$ & $5176.05$ & $[$Fe $\textsc{vii}]$ & $5276.38$ & O $\textsc{vi}$ & $5291.00$ \\
$[$Fe $\textsc{xiv}]$ & $5303.01$ & $[$Ca $\textsc{v}]$ & $5309.11$ & He $\textsc{ii}$ & $5411.37$ & $[$Fe $\textsc{vi}]$ & $5424.23$ & $[$Ar $\textsc{x}]$ & $5534.02$ \\
$[$Ca $\textsc{vii}]$ & $5618.76$ & $[$Fe $\textsc{vi}]$ & $5631.08$ & $[$Fe $\textsc{vi}]$ & $5676.96$ & N $\textsc{ii}$ & $5679.00$ & $[$Fe $\textsc{vii}]$ & $5720.71$ \\
He $\textsc{i}$ & $5875.64$ & $[$Ca $\textsc{v}]$ & $6086.37$ & $[$Fe $\textsc{vii}]$ & $6086.97$ & $[$Fe $\textsc{x}]$ & $6374.54$ & $[$Ar $\textsc{v}]$ & $6435.12$ \\
He $\textsc{ii}$ & $6559.91$ & H $\textsc{i}$ & $6562.81$ & $[$N $\textsc{ii}]$ & $6583.45$ & He $\textsc{i}$ & $6678.15$ & He $\textsc{ii}$ & $6682.98$ \\
He $\textsc{ii}$ & $6890.67$ & $[$Ar $\textsc{v}]$ & $7005.83$ & He $\textsc{i}$ & $7065.22$ & He $\textsc{ii}$ & $7177.28$ & N $\textsc{iv}$ & $7582.00$ \\
He $\textsc{ii}$ & $7592.50$ & N $\textsc{iv}$ & $7703.00$ & O $\textsc{iv}$ & $7713.00$ & $[$Fe $\textsc{xi}]$ & $7891.87$ & $[$Mn $\textsc{ix}]$ & $7968.49$ \\
He $\textsc{ii}$ & $8236.51$ & H $\textsc{i}$ & $8750.43$ & H $\textsc{i}$ & $8862.74$ & H $\textsc{i}$ & $9014.87$ & $[$S $\textsc{iii}]$ & $9068.62$ \\
H $\textsc{i}$ & $9228.97$ & He $\textsc{ii}$ & $9344.62$ & $[$S $\textsc{iii}]$ & $9530.62$ & H $\textsc{i}$ & $9545.93$ & H $\textsc{i}$ & $10049.30$ \\
He $\textsc{ii}$ & $10123.30$ & He $\textsc{i}$ & $10830.30$ & H $\textsc{i}$ & $10938.00$ & H $\textsc{i}$ & $12818.00$ & H $\textsc{i}$ & $18751.00$ \\
H $\textsc{i}$ & $19445.40$ & $[$Al $\textsc{ix}]$ & $20444.40$ & H $\textsc{i}$ & $21655.10$ & $[$Ca $\textsc{viii}]$ & $23211.70$ & $[$Si $\textsc{vii}]$ & $24807.10$ \\
\enddata
\end{deluxetable*}

\begin{deluxetable*}{cccccccccc}
\tablecaption{Epoch-Selection for 2560 days age.}
\label{tab:Epoch-Selection_2560d}
\tablehead{
\colhead{Element} & \colhead{$\lambda$ ($\lambda$\AA)} & 
\colhead{Element} & \colhead{$\lambda$ ($\lambda$\AA)} & 
\colhead{Element} & \colhead{$\lambda$ ($\lambda$\AA)} & 
\colhead{Element} & \colhead{$\lambda$ ($\lambda$\AA)} & 
\colhead{Element} & \colhead{$\lambda$ ($\lambda$\AA)}
}
\startdata
He $\textsc{ii}$ & $3203.04$ & $[$Ne $\textsc{v}]$ & $3345.99$ & $[$Ne $\textsc{v}]$ & $3426.03$ & O $\textsc{vi}$ & $3434.00$ & $[$Fe $\textsc{vi}]$ & $3492.10$ \\
$[$Fe $\textsc{vi}]$ & $3555.61$ & $[$Fe $\textsc{vii}]$ & $3586.32$ & $[$Fe $\textsc{vi}]$ & $3662.50$ & H $\textsc{i}$ & $3734.36$ & H $\textsc{i}$ & $3750.15$ \\
$[$Fe $\textsc{vii}]$ & $3758.92$ & H $\textsc{i}$ & $3770.63$ & H $\textsc{i}$ & $3797.89$ & $[$Fe $\textsc{vi}]$ & $3813.55$ & H $\textsc{i}$ & $3835.38$ \\
He $\textsc{ii}$ & $3857.96$ & $[$Ne $\textsc{iii}]$ & $3868.76$ & He $\textsc{i}$ & $3888.63$ & He $\textsc{ii}$ & $3923.37$ & $[$Ne $\textsc{iii}]$ & $3967.47$ \\
H $\textsc{i}$ & $3970.07$ & $[$Fe $\textsc{xi}]$ & $3986.88$ & He $\textsc{i}$ & $4026.20$ & $[$Fe $\textsc{v}]$ & $4071.24$ & H $\textsc{i}$ & $4101.73$ \\
$[$Fe $\textsc{v}]$ & $4180.59$ & C $\textsc{iii}$ & $4187.00$ & H $\textsc{i}$ & $4340.46$ & $[$O $\textsc{iii}]$ & $4363.21$ & He $\textsc{i}$ & $4471.49$ \\
He $\textsc{ii}$ & $4541.46$ & N $\textsc{iii}$ & $4641.00$ & C $\textsc{iv}$ & $4659.00$ & He $\textsc{ii}$ & $4685.64$ & $[$Ar $\textsc{iv}]$ & $4740.12$ \\
Fe $\textsc{vii}$ & $4893.37$ & $[$Ca $\textsc{vii}]$ & $4939.56$ & $[$Fe $\textsc{vii}]$ & $4942.48$ & N $\textsc{v}$ & $4945.00$ & $[$O $\textsc{iii}]$ & $4958.91$ \\
$[$Fe $\textsc{vi}]$ & $4967.15$ & $[$Fe $\textsc{vii}]$ & $4988.55$ & $[$O $\textsc{iii}]$ & $5006.84$ & $[$Fe $\textsc{vi}]$ & $5145.76$ & $[$Fe $\textsc{vii}]$ & $5158.89$ \\
$[$Fe $\textsc{vi}]$ & $5176.05$ & $[$Fe $\textsc{vii}]$ & $5276.38$ & O $\textsc{vi}$ & $5291.00$ & $[$Fe $\textsc{xiv}]$ & $5303.01$ & $[$Ca $\textsc{v}]$ & $5309.11$ \\
He $\textsc{ii}$ & $5411.37$ & $[$Fe $\textsc{vi}]$ & $5424.23$ & $[$Ar $\textsc{x}]$ & $5534.02$ & $[$Ca $\textsc{vii}]$ & $5618.76$ & $[$Fe $\textsc{vi}]$ & $5631.08$ \\
$[$Fe $\textsc{vi}]$ & $5676.96$ & $[$Fe $\textsc{vii}]$ & $5720.71$ & He $\textsc{i}$ & $5875.64$ & $[$Ca $\textsc{v}]$ & $6086.37$ & $[$Fe $\textsc{vii}]$ & $6086.97$ \\
$[$Fe $\textsc{x}]$ & $6374.54$ & $[$Ar $\textsc{v}]$ & $6435.12$ & He $\textsc{ii}$ & $6559.91$ & H $\textsc{i}$ & $6562.81$ & He $\textsc{i}$ & $6678.15$ \\
He $\textsc{ii}$ & $6682.98$ & He $\textsc{ii}$ & $6890.67$ & $[$Ar $\textsc{v}]$ & $7005.83$ & He $\textsc{i}$ & $7065.22$ & He $\textsc{ii}$ & $7177.28$ \\
N $\textsc{iv}$ & $7582.00$ & He $\textsc{ii}$ & $7592.50$ & N $\textsc{iv}$ & $7703.00$ & O $\textsc{iv}$ & $7713.00$ & $[$Fe $\textsc{xi}]$ & $7891.87$ \\
$[$Mn $\textsc{ix}]$ & $7968.49$ & He $\textsc{ii}$ & $8236.51$ & H $\textsc{i}$ & $8545.34$ & H $\textsc{i}$ & $8598.35$ & H $\textsc{i}$ & $8664.98$ \\
H $\textsc{i}$ & $8750.43$ & H $\textsc{i}$ & $8862.74$ & H $\textsc{i}$ & $9014.87$ & H $\textsc{i}$ & $9228.97$ & He $\textsc{ii}$ & $9344.62$ \\
$[$S $\textsc{iii}]$ & $9530.62$ & H $\textsc{i}$ & $9545.93$ & H $\textsc{i}$ & $10049.30$ & He $\textsc{ii}$ & $10123.30$ & He $\textsc{i}$ & $10830.30$ \\
H $\textsc{i}$ & $10938.00$ & H $\textsc{i}$ & $12818.00$ & H $\textsc{i}$ & $18174.00$ & H $\textsc{i}$ & $18751.00$ & H $\textsc{i}$ & $19445.40$ \\
$[$Al $\textsc{ix}]$ & $20444.40$ & H $\textsc{i}$ & $21655.10$ & He $\textsc{ii}$ & $21884.30$ & $[$Ca $\textsc{viii}]$ & $23211.70$ & $[$Si $\textsc{vii}]$ & $24807.10$ \\
\enddata
\end{deluxetable*}

\newpage

\bibliography{Main}{}
\bibliographystyle{aasjournal}

\end{document}